\documentclass[english,notitlepage,nofootinbib]{revtex4-1}
\pdfoutput=1
\usepackage[T1]{fontenc}
\usepackage[latin9]{inputenc}
\usepackage{color}
\usepackage{textcomp}
\usepackage{amsmath}
\usepackage{amssymb}
\usepackage{stmaryrd}
\usepackage{graphicx}
\usepackage{hyperref}

\makeatletter

\providecommand{\tabularnewline}{\\}

\usepackage{slashed}
\usepackage[percent]{overpic}

\makeatother

\usepackage{babel}

\hypersetup{colorlinks=true, linkcolor={black}, citecolor={black}, urlcolor={black}}

\begin{document}

\title{Loop-induced Neutrino Non-Standard Interactions }

\author{ Ingolf Bischer, Werner Rodejohann, Xun-Jie Xu}

\affiliation{\textcolor{black}{Max-Planck-Institut f\"ur Kernphysik, Postfach
103980, D-69029 Heidelberg, Germany}.}

\begin{abstract}
\noindent
Non-Standard Interactions (NSI) of neutrinos may originate from models in which new 
particles interact with neutrinos. In scalar extensions of the SM, the 
typical approach to obtain NSI requires Fierz transformations and 
charged Higgses, which suffer from strong constraints from collider 
searches or charged lepton flavor violation processes. We propose here 
an alternative approach to generate NSI, namely via loop processes. We 
show that such loop-induced NSI from secret neutrino interactions can 
reach sizes of ${\cal O}(0.1\sim1)$ compared to standard Fermi interaction. 
This approach can also give rise to neutrino-quark NSI. 
\end{abstract}
\maketitle

\section{Introduction}

In the era of neutrino oscillation precision measurements, the standard
three-neutrino oscillation framework is being tested with increasing
precision \cite{Capozzi:2016rtj,Esteban:2016qun,deSalas:2017kay}.
 Hence it is important to consider any new physics that could have
significant effects on neutrino oscillations. One particularly interesting 
possibility is provided by so-called Non-Standard Interactions (NSI) of neutrinos, 
which has raised rather general interest in the literature (see, 
e.g., Refs.\ \cite{Davidson:2003ha,Ohlsson:2012kf,Farzan:2017xzy,Esteban:2018ppq} for reviews on NSI). 
 By introducing new flavor-changing neutral-current interactions 
($\overline{\nu_{\alpha}}\gamma^{\mu}\nu_{\beta}\overline{\psi}\gamma_{\mu}\psi$)
of neutrinos ($\nu_{\alpha}$, $\nu_{\beta}$) with other Standard Model fermions 
$\psi$, such NSI cause via coherent forward scattering flavor transitions in matter, disturbing the determination of 
the standard neutrino physics parameters. 
The effects of NSI in current and future long-baseline
experiments (T2K, NO$\nu$A, DUNE, etc.), especially on the determination 
of $\delta_{CP}$, have been 
extensively studied \cite{Masud:2015xva,deGouvea:2015ndi,Masud:2016bvp,Masud:2016gcl,Blennow:2016etl,Agarwalla:2016fkh,Deepthi:2017gxg}. 

From the theoretical point of view, NSI of neutrinos are 
well motivated. Generally speaking, neutrinos have long been considered
as the portal of new physics, even more so after they were found to be massive. It is 
reasonable to speculate that the new physics related to neutrinos
also brings new interactions to neutrinos. A well-known example is 
the type II seesaw model \cite{Konetschny:1977bn,Cheng:1980qt,Schechter:1980gr}.
In this model, a scalar triplet is introduced to the SM and acquires 
a small vacuum exception value to generate neutrino
masses\footnote{See \cite{Xu:2016klg} for a recent analysis on how to achieve this.}.
Since the triplet couples both to electrons and neutrinos, 
NSI of neutrinos with electrons can be generated \cite{Malinsky:2008qn}. 
In addition to the type II seesaw model, other scalar extensions
of the SM can also generate NSI in the same way, including scalar
singlet models \cite{Bilenky:1993bt,Antusch:2008tz,Wise:2014oea,Forero:2016ghr} or two-Higgs-doublet
models \cite{Herrero-Garcia:2017xdu,Dey:2018yht}, etc. In all these
scalar extensions, the approach of generating NSI is to integrate
out a  charged scalar mediator to get scalar four-fermion interactions which are 
then converted by a Fierz transformation 
to vector form (containing $\gamma^{\mu}$). The mediator 
must be charged due to the Fierz transformation rules (as we will demonstrate explicitly later), 
which is potentially a problem of obtaining
sizable NSI because charged Higgses usually face stronger collider
constraints than neutral ones. 

In this paper, we propose a different way to generate NSI, namely
loop-induced NSI. The approach is also based on scalar extensions of the 
SM\footnote{Gauge extensions may also generate NSI of neutrinos, by 
integrating out a flavor-sensitive $Z'$, e.g., in gauged 
$L_{\mu}-L_{\tau}$ models \cite{Heeck:2011wj}. One can also imagine 
scenarios in which $Z'$ models generate NSI via loops. Here we focus on the scalar case, 
 since the scalar sector
is the least experimentally tested, leaving a larger parameter
space unexplored.}, but without using Fierz transformations. 
Instead, as the name
implies, the loop-induced NSI are generated by loop diagrams. Although 
loop contributions are in general expected to be subdominant compared
with tree level contributions, in some models this way can produce
fairly sizable NSI which is absent at tree level. The advantage 
of loop-induced NSI compared to the usual one obtained by the Fierz
transformation and charged Higgses is that the source of flavor violation
can be confined to the neutrino sector with ``hidden'' scalar interactions. 
Hence, large NSI can be obtained without causing problems in other well-measured 
processes. Other scenarios can also give rise to neutrino-quark NSI, which are 
absent in the previous models.

The remainder of this paper is organized as follows. In Sec.~\ref{sec:General-analyses},
we first briefly review how NSI can be generated in scalar models by Fierz transformations, 
and then introduce our concept of generating NSI by loop diagrams,
with some general results presented while the detailed calculation
is delegated to the appendices. Then we apply the results to several 
explicit models in Sec.~\ref{sec:scalar-singlet} to \ref{sec:Model-C}. 
Confronting these models with experimental constraints, we estimate
the order of magnitude of the loop-induced NSI in these models in
Sec.~\ref{sec:How-large}. Finally we conclude in Sec.~\ref{sec:Conclusion}.

\section{General analysis\label{sec:General-analyses}}

In this section, we study the generation of NSI in a general framework
which introduces a new scalar boson $\phi$. It has Yukawa interactions
with neutrinos and probably other SM fermions.  Let us consider how
the following NSI may be generated,
\begin{equation}
{\cal L}_{{\rm NSI}}=
\frac{G_{F}}{\sqrt{2}}\epsilon_{\alpha\beta}^{\psi}\overline{\psi}\gamma^{\mu}\psi\overline{\nu_{\alpha}}\gamma_{\mu}P_{L}\nu_{\beta},
\label{eq:L-48}
\end{equation}
Here $P_{L}\equiv(1-\gamma^{5})/2$ and $\psi$ stands for electrons
or quarks which can be chiral (e.g.\ $\psi=e_{L}$, $u_{R}$, $d_{L}, \cdots$). Throughout the paper, we use $\alpha$, $\beta, \cdots$ 
to denote the flavor indices. 

In practice, NSI are usually expressed in terms of non-chiral neutrons ($n$), protons ($p$) and electrons ($e$):
\begin{equation}
{\cal L}_{{\rm NSI}}=
\frac{G_{F}}{\sqrt{2}}
\overline{\nu_{\alpha}}\gamma_{\mu}P_{L}\nu_{\beta}
\left[\overline{e}\gamma^{\mu}\left(\epsilon_{\alpha\beta}^{e,V}+\epsilon_{\alpha\beta}^{e,A}\gamma^{5}\right)e+\overline{n}\gamma^{\mu}\left(\epsilon_{\alpha\beta}^{n,V}+\epsilon_{\alpha\beta}^{n,A}\gamma^{5}\right)n+\overline{p}\gamma^{\mu}\left(\epsilon_{\alpha\beta}^{p,V}+\epsilon_{\alpha\beta}^{p,A}\gamma^{5}\right)p\right]
.
\label{eq:L-48-explicit}
\end{equation}
The NSI couplings in Eq.~(\ref{eq:L-48-explicit}) can be connected to the chiral form in (\ref{eq:L-48}) by\footnote{
Axial NSI of nucleons or electrons are not important in neutrino oscillations, we hence ignore this part in this paper.
} 
\begin{equation}
\epsilon_{\alpha\beta}^{e,V}=\epsilon_{\alpha\beta}^{e_{L}}+\epsilon_{\alpha\beta}^{e_{R}},\ \ \epsilon_{\alpha\beta}^{e,A}=\epsilon_{\alpha\beta}^{e_{R}}-\epsilon_{\alpha\beta}^{e_{L}}
,\label{eq:VALRe}
\end{equation}
\begin{equation}
\epsilon_{\alpha\beta}^{n,V}=(\epsilon_{\alpha\beta}^{u_{L}}+\epsilon_{\alpha\beta}^{u_{R}})+2(\epsilon_{\alpha\beta}^{d_{L}}+\epsilon_{\alpha\beta}^{d_{R}}),\ \ \epsilon_{\alpha\beta}^{p,V}=2(\epsilon_{\alpha\beta}^{u_{L}}+\epsilon_{\alpha\beta}^{u_{R}})+(\epsilon_{\alpha\beta}^{d_{L}}+\epsilon_{\alpha\beta}^{d_{R}})
.\label{eq:VALRnp}
\end{equation}
Currently the experimental constraints on these NSI parameters, depending on the specific channels, range from ${\cal O}(10^{-2})$ to ${\cal O}(1)$---for a recent update, see Ref.~\cite{Esteban:2018ppq}.

To obtain the operator in Eq.~(\ref{eq:L-48}), we need two essentials: 
one is flavor-sensitive interactions 
of the new scalar boson and the other is the conversion
of the scalar form\footnote{In this paper, we refer to fermion interactions with the Dirac matrices
$\mathbf{1}$, $\gamma^{5}$, $\gamma^{\mu}$, $\gamma^{\mu}\gamma^{5}$,
and $\sigma^{\mu\nu}$ between the fermion fields as scalar, pseudo-scalar,
vector, axial-vector, and tensor forms, respectively. For example,
$\overline{\psi}\gamma^{\mu}\psi A_{\mu}$ and $\overline{\psi}\gamma^{\mu}\psi\overline{\psi}\gamma^{\mu}\psi$
are vector form interactions; $\overline{\psi}\psi\phi$ and $\overline{\psi}\psi\overline{\psi}\psi$
are scalar form interactions.} to vector form. More technically, the NSI operators
contain $\gamma^{\mu}$ while the new scalar boson only introduces 
interactions which do not contain $\gamma^{\mu}$. 
Here we introduce two approaches to achieve the conversion, by the
Fierz transformation and by loop corrections. We will refer to the
corresponding  NSI as Fierz-transformed NSI and loop-induced NSI respectively.

\subsection{Fierz-transformed NSI}

Applying the Fierz transformations in some scalar extensions of the
SM (e.g.\ the type II seesaw model) to obtain NSI has been considered
in the literature \cite{Malinsky:2008qn,Antusch:2008tz,Ohlsson:2009vk,Herrero-Garcia:2017xdu}. 
Generally, if a heavy scalar boson $\phi$ has Yukawa interactions
$\overline{\psi_{1}}\psi_{2}\phi$ and $\overline{\psi_{3}}\psi_{4}\phi$,
integrating it out will lead to the four-fermion effective 
operator $\overline{\psi_{1}}\psi_{2}\overline{\psi_{3}}\psi_{4}$.
The Fierz transformation (see, e.g., \cite{Giunti})  of this operator
gives
\begin{align}
\overline{\psi_{1}}\psi_{2}\overline{\psi_{3}}\psi_{4} & =-\frac{1}{4}\overline{\psi_{1}}\psi_{4}\overline{\psi_{3}}\psi_{2}-\frac{1}{4}\overline{\psi_{1}}\gamma^{5}\psi_{4}\overline{\psi_{3}}\gamma^{5}\psi_{2}\nonumber \\
 & -\frac{1}{4}\overline{\psi_{1}}\gamma^{\mu}\psi_{4}\overline{\psi_{3}}\gamma_{\mu}\psi_{2}+\frac{1}{4}\overline{\psi_{1}}\gamma^{\mu}\gamma^{5}\psi_{4}\overline{\psi_{3}}\gamma_{\mu}\gamma^{5}\psi_{2}\nonumber \\
 & -\frac{1}{8}\overline{\psi_{1}}\sigma^{\mu\nu}\psi_{4}\overline{\psi_{3}}\sigma_{\mu\nu}\psi_{2},\label{eq:L-49}
\end{align}
where the third term on the right-hand-side 
is a vector form interaction. Recall that only the vector form interaction 
leads to NSI effects in terrestrial matter \cite{Bergmann:1999rz}. 
In the SM and many
extensions, the Yukawa interactions are based on chiral fermions.
So it is also useful to provide the Fierz transformations of chiral
fermions:
\begin{align}
\overline{\psi_{1}}P_{L}\psi_{2}\overline{\psi_{3}}P_{L}\psi_{4} & =\overline{\psi_{1R}}\psi_{2L}\overline{\psi_{3R}}\psi_{4L}=-\frac{1}{2}\overline{\psi_{1}}P_{L}\psi_{4}\overline{\psi_{3}}P_{L}\psi_{2}-\frac{1}{8}\overline{\psi_{1}}\sigma^{\mu\nu}P_{L}\psi_{4}\overline{\psi_{3}}\sigma_{\mu\nu}P_{L}\psi_{2},\label{eq:L-50}\\
\overline{\psi_{1}}P_{L}\psi_{2}\overline{\psi_{3}}P_{R}\psi_{4} & =\overline{\psi_{1R}}\psi_{2L}\overline{\psi_{3L}}\psi_{4R}=-\frac{1}{2}\overline{\psi_{1}}\gamma^{\mu}P_{R}\psi_{4}\overline{\psi_{3}}\gamma_{\mu}P_{L}\psi_{2},\label{eq:L-51}
\end{align}
which can be obtained by replacing $(\overline{\psi_{1}},\ \psi_{2},\ \overline{\psi_{3}},\ \psi_{4})$
in Eq.~(\ref{eq:L-49}) with $(\overline{\psi_{1R}},\ \psi_{2L},\ \overline{\psi_{3R}},\ \psi_{4L})$
and $(\overline{\psi_{1R}},\ \psi_{2L},\ \overline{\psi_{3L}},\ \psi_{4R})$.
It is noteworthy that Eq.~(\ref{eq:L-50}) produces only scalar and
tensor form interactions while Eq.~(\ref{eq:L-51}) produces only
vector form interactions. Therefore, in a chiral theory only when
the effective operator has a chirality structure as in Eq.~(\ref{eq:L-51}),
the vector form NSI can be obtained. 

If Eq.~(\ref{eq:L-49}) or Eq.~(\ref{eq:L-51}) is used to generate
NSI, we should identify $\psi_{2}$ and $\overline{\psi_{3}}$ with 
neutrinos, and $\psi_{4}$ and $\overline{\psi_{1}}$ 
with electrons or quarks. Note that $\psi_{1}$ and $\psi_{4}$ need to be identical to 
generate NSI terms from coherent forward scattering in matter. 
This is clear from 
comparing Eqs.\ (\ref{eq:L-49}), (\ref{eq:L-51}) with Eq.~(\ref{eq:L-48}). 
In addition, limits from flavor physics strongly contrains cases with $\psi_{1} \neq\psi_{4}$. 
Hence we can infer that the effective operator before the Fierz transformation should
be $\overline{\psi}\nu_{L}\overline{\nu_{L}}\psi$, where $\psi$ 
stands for charged fermions. 
Since $\overline{\psi}\nu_{L}$ and $\overline{\nu_{L}}\psi$
have nonzero electric charges, the new scalar boson must be charged.
If $\psi$ is a quark, then the scalar boson has to be colored.
Such leptoquarks are severely constrained. 

In conclusion, the Fierz transformation approach requires a charged 
scalar boson to generate NSI. If the boson is a singlet under $SU(3)_{c}$,
then neutrino-quark NSI can not be generated. Note further that since strong limits on additional charged scalars exist, the particle responsible for the 
Fierz-transformed NSI can not be light (MeV-scale), which is often discussed (see e.g.\ 
\cite{Farzan:2015doa,Farzan:2016wym})  
in the context of matter-induced NSI by coherent forward scattering. 

\subsection{Loop-induced NSI\label{subsec:Loop-induced-NSI}}

\begin{figure}
\centering

\begin{overpic}[height=4.0cm]
{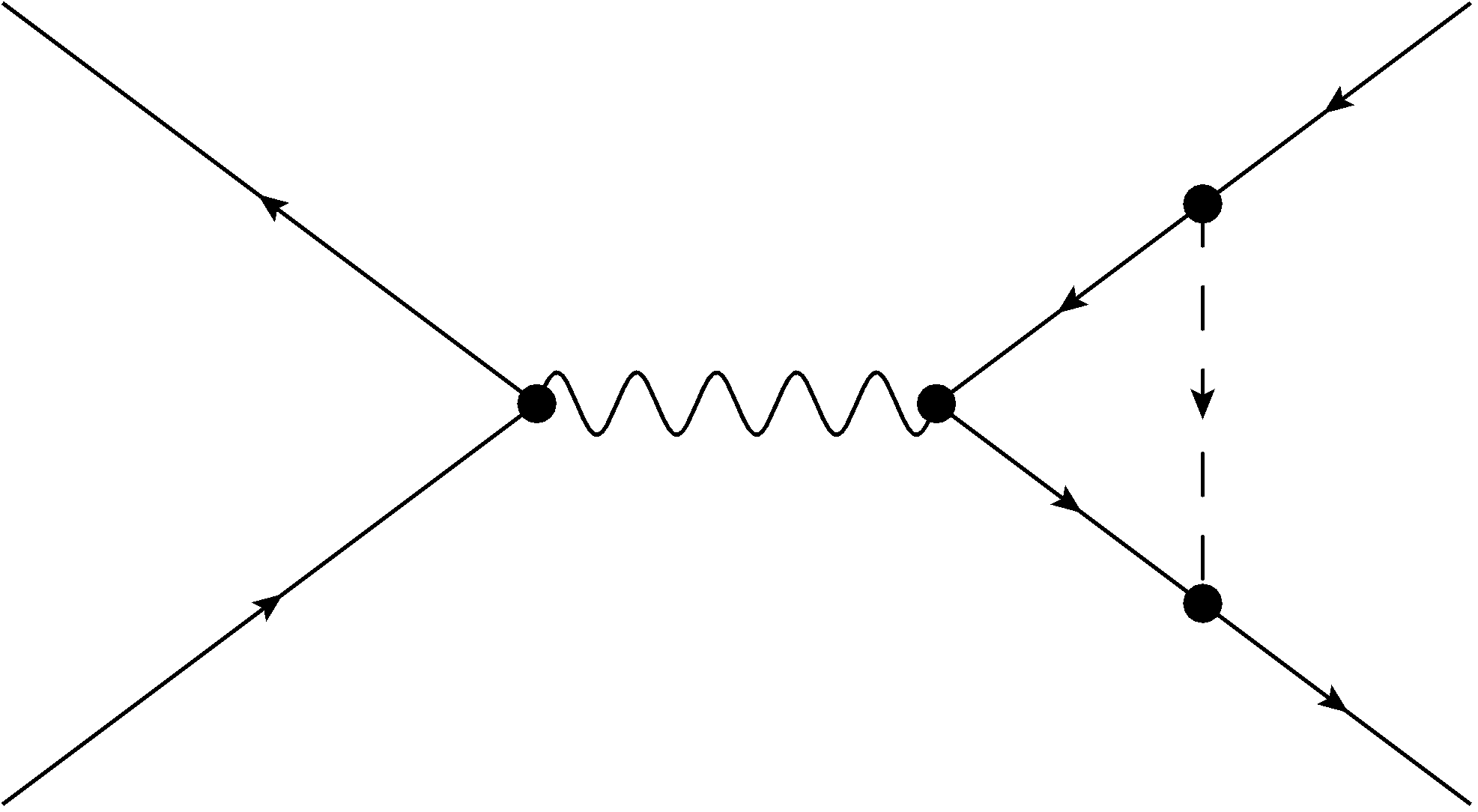} 
\put (50,33) {$Z_{\mu}$}
\put (10,50) {$\psi=e$, $u$, $d$}
\put (10,5) {$\psi=e$, $u$, $d$}
\put (92,46) {$\nu_{\beta}$} \put (92,9) {$\nu_{\alpha}$}
\put (68,39) {$\psi_{{\rm int}}$}
\put (68,15) {$\psi_{{\rm int}}$}
\put (82,30) {$\phi$}

\end{overpic}\hspace{1cm}
\begin{overpic}[height=4.0cm]
{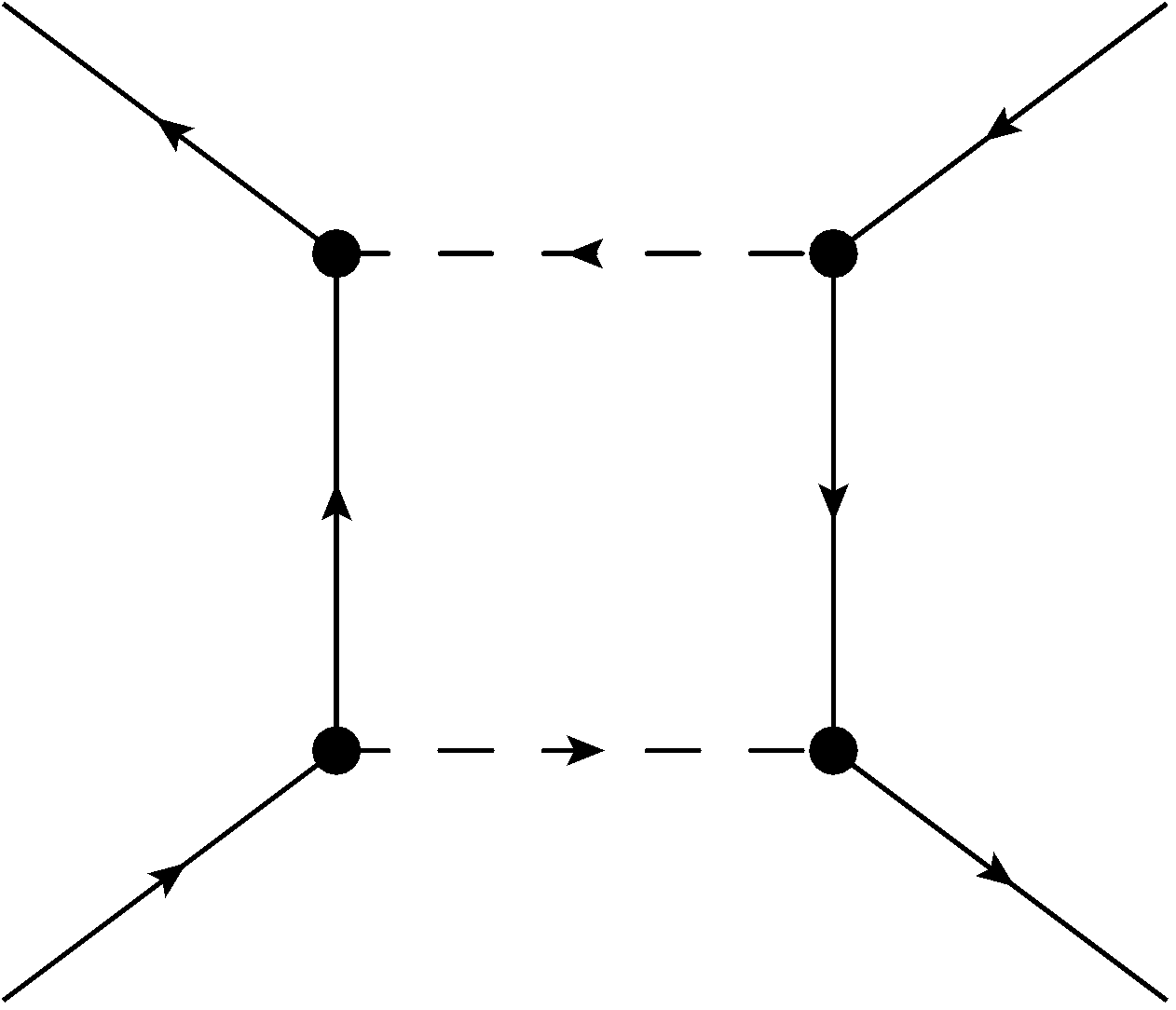} 
\put (8,80) {$\psi=e$, $u$, $d$} \put (8,3) {$\psi=e$, $u$, $d$}
\put (92,75) {$\nu_{\beta}$} \put (92,10) {$\nu_{\alpha}$}
\put (31,39) {$\psi'$}
\put (20,60) {$y_{\psi}^*$}
\put (20,25) {$y_{\psi}$}
\put (50,70) {$\phi$} \put (50,26) {$\phi$}

\end{overpic}\hspace{1cm}

\caption{\label{fig:NSI-loop}Triangle and box diagrams which generate the
NSI in Eq.~(\ref{eq:L-65}). }
\end{figure}

We will demonstrate now that if neutrinos have Yukawa interactions with a new scalar boson, then
NSI can be generated at the loop level\footnote{Note that in the SM loop-induced and flavor-diagonal NSI are present. Their magnitude can be estimated to be of order $\epsilon \sim 
m_\tau^2/(16 \pi^2 m_W^2) \sim 10^{-6}$, hence completely negligible. }. Both neutral and charged 
Higgses can generate such terms. Here we discuss two possible
diagrams for loop-induced NSI, as shown in Fig.~\ref{fig:NSI-loop}.
The first one is based on loop corrections to the neutrino-$Z$ vertex
(left panel) which we will refer to as the triangle diagram. The other
is a box diagram, which consists of pure Yukawa interactions and does
not involve any gauge interactions. The external fermion lines are
two neutrinos of different flavor, and two charged fermions, which
can be either electrons or quarks. 
The internal fermion lines can be charged or neutral fermions and do not need to be identical,  depending on the models. 
As discussed above, the two external charged fermions should be identical. 

As we have mentioned, the flavor violation is introduced by the scalar-neutrino
interactions and needs to be converted to vector form interactions. 
In the triangle diagram, this is achieved by the fact that the triangle 
loop generates an effective flavor-changing vertex $Z_{\mu}\overline{\nu_{L\alpha}}\gamma^{\mu}\nu_{L\beta}$.
In the box diagram, the effective four-fermion operator also has $\gamma^{\mu}$'s
between the fermion fields because of the internal fermion propagator. 

In computing the loop-induced NSI, we need to consider the UV divergences.
By simple power counting, one can see that the triangle diagram contains
a logarithmic UV divergence $\int^{\Lambda}d^{4}k\frac{1}{k^{4}}\sim \log \Lambda$
while the box diagram is not divergent because $\int d^{4}k\frac{1}{k^{6}}$
is finite. In a renormalizable model, the UV divergence in any physical
process should be canceled by adding all relevant diagrams and
counterterms together. For the triangle diagram considered here, 
because at tree level the neutral current interactions are flavor 
conserving, there is no corresponding counter term. Therefore in a
renormalizable and complete model, one simply needs to sum over the
relevant diagrams to obtain a finite result. In one of the models 
considered below, the cancellation of divergences 
is ensured by the conservation of the gauge charges.

\begin{figure}
\centering

\includegraphics[height=5cm]{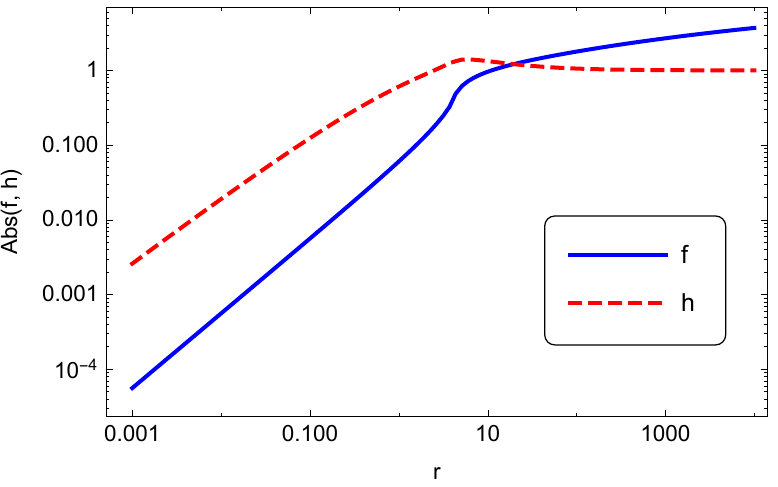}\ \ \includegraphics[height=5cm]{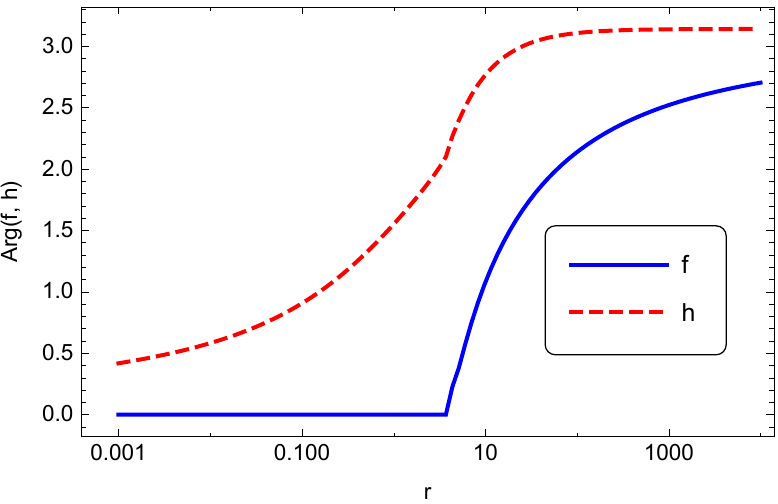}

\caption{\label{fig:hg}Numerical values of the functions $f(r)$ and $h(r)$ 
in Eq.~(\ref{eq:L-55}) with  $r\equiv m_{Z}^{2}/m_{\phi}^{2}$.
}
\end{figure}

\newpage

\vspace{0.3cm}
\noindent{\bf Triangle diagrams:}\newline
In Appendix~\ref{sec:triangle-diagrams}, we compute the triangle 
diagram and the result is presented as follows. If the UV divergences
cancel out, the effective flavor changing $Z-\nu$ vertex in Fig.~\ref{fig:NSI-loop}
is
\begin{equation}
{\cal L}_{{\rm eff}}=g_{\alpha\beta}^{(1)}Z_{\mu}\overline{\nu_{\alpha}}\gamma^{\mu}P_{L}\nu_{\beta},\ {\rm (no\ sum\ over\ \alpha,\ \beta)},
\mbox{ with } 
g_{\alpha\beta}^{(1)}=\frac{y_{\alpha}^{*}y_{\beta}}{16\pi^{2}}\frac{g}{c_{W}}\frac{m_{Z}^{2}}{m_{\phi}^{2}}\left[f(r)Q_{Z}^{(\nu_{L})}+h(r)Q_{Z}^{(\psi_{{\rm int}})}\right].\label{eq:L-55}
\end{equation}
Note that this result is derived under the assumption that the masses of the fermions involved are all negligibly small, which implies that in the limit $m_{\phi}\rightarrow 0$, Eq.~(\ref{eq:L-55}) does not give a valid result.
The notations in Eq.~(\ref{eq:L-55}) are explained as follows:
\begin{itemize}
\item $\psi_{{\rm int}}$ is the internal fermion appearing in the triangle loop.
The Yukawa vertices are formulated as
\begin{equation}
{\cal L}\supset y_{\alpha}\overline{\psi}_{{\rm int}}\phi\nu_{L\alpha}+y_{\beta}\overline{\psi}_{{\rm int}}\phi\nu_{L\beta}+{\rm h.c.},\label{eq:L-53}
\end{equation}
which defines the Yukawa couplings $y_{\alpha}$ and $y_{\beta}$.
In the triangle diagram, $\psi_{{\rm int}}$ can be any SM fermion that 
couples to the $Z$ boson. In the loop calculation, we assume the fermion masses are
all negligibly small compared to the boson masses (both $\phi$ and
$Z$). This is fine as long as only leptons are coupling to the scalar, but one could also 
accomodate more exotic models where quarks including the top couple to scalars and neutrinos. 
\item $g/c_{W}$ is the gauge coupling attached to the $Z$ boson, and $Q_{Z}$
is the corresponding $Z$ charge of a fermion. Both are defined by the covariant
derivative
\begin{equation}
D_{\mu}=\partial_{\mu}-i\frac{g}{c_{W}}Z_{\mu}Q_{Z}.\label{eq:L-56}
\end{equation}
For convenience we list the $Z$ charges of the SM fermions 
  in Tab.~\ref{tab:Z-charge}.
\end{itemize}
\begin{table*}
\caption{\label{tab:Z-charge}$Z$ charges of Standard Model  fermions. }
\begin{ruledtabular}
\begin{minipage}{0.8\textwidth}

\begin{tabular}{ccccccccccc}
 & $\nu_{L}$ & $e_{L}$ & $e_{R}$ & $u_{L}$ & $u_{R}$ & $d_{L}$ & $d_{R}$ & $e$ & $n$ & $p$\tabularnewline
\hline 
$Q_{Z}$ & $\frac{1}{2}$ & $-\frac{1}{2}+s_{W}^{2}$ & $s_{W}^{2}$ & $\frac{1}{2}-\frac{2}{3}s_{W}^{2}$ & $-\frac{2}{3}s_{W}^{2}$ & $-\frac{1}{2}+\frac{1}{3}s_{W}^{2}$ & $\frac{1}{3}s_{W}^{2}$ & $-\frac{1}{2}+2s_{W}^{2}$ & $-\frac{1}{2}$ & $\frac{1}{2}-2s_{W}^{2}$\tabularnewline
\end{tabular}

\end{minipage}
\end{ruledtabular}

\end{table*}

\begin{itemize}
\item The scalar boson $\phi$ in the triangle loop has mass $m_{\phi}$.
 Depending on models, it may also have a $Z$-charge $Q_{Z}^{(\phi)}$.
In Appendix~\ref{sec:triangle-diagrams}, we show that if the $Z$
charges are conserved in the model, then the UV divergences 
cancel. In Eq.~(\ref{eq:L-53}), the $Z$ charge conservation
requires
\begin{equation}
Q_{Z}^{(\phi)}=Q_{Z}^{(\psi_{{\rm int}})}-Q_{Z}^{(\nu_{L})}.\label{eq:L-57}
\end{equation}
Any renormalizable model satisfies
Eq.~(\ref{eq:L-57}), as we demonstrate explicitely in 
In Sec.~\ref{sec:scalar-singlet}. 
\item $f(r)$ and $h(r)$  are two finite functions of the mass ratio 
\begin{equation}
r\equiv m_{Z}^{2}/m_{\phi}^{2}.\label{eq:L-58}
\end{equation}
The explicit forms of $f(r)$ and $h(r)$ are rather complicated and can be found 
in Appendix~\ref{sec:triangle-diagrams}. These functions 
have simple limits for $r\gg1$ and $r\ll1$:
\begin{align}
r & \gg1:\ \ f(r)\approx\frac{5}{4}-\frac{\log r}{2}+\frac{\pi}{2}i,\ \ h(r)\approx-1-\frac{\log r}{r},\label{eq:L-59}\\
r & \ll1:\ \ f(r)\approx\frac{r}{18},\ \ h(r)\approx\frac{r}{18}(1-6\log r+6i\pi).\label{eq:L-60}
\end{align}
 For general values of $r$, we numerically evaluate them and show 
the results in Fig.~\ref{fig:hg}.
\end{itemize}

Given the effective $Z\overline{\nu}\nu$ vertex in Eq.~(\ref{eq:L-55}),
the corresponding low-energy four-fermion interaction is 
\begin{equation}
{\cal L}_{{\rm NSI}}^{\triangleright}=-\frac{G_{F}}{\sqrt{2}}\frac{8g_{\alpha\beta}^{(1)}}{g}Q_{Z}^{(\psi)}c_{W}\overline{\psi}\gamma^{\mu}\psi\overline{\nu_{\alpha}}\gamma_{\mu}P_{L}\nu_{\beta},\label{eq:L-63}
\end{equation}
where $G_{F}=\sqrt{2}g^{2}/(8m_{Z}^{2}c_{W}^{2})$. 

\vspace{0.3cm}
\noindent{\bf Box diagrams:}\newline
The box diagram is always finite. After computing the loop integral
(see Appendix~\ref{sec:box-diagram}), we obtain the effective Lagrangian generated by the 
box NSI:
\begin{equation}
{\cal L}_{{\rm NSI}}^{\boxempty}=\frac{1}{16\pi^{2}}\frac{y_{\alpha}^{*}y_{\beta}|y_{\psi}|^{2}}{4m_{\phi}^{2}}\overline{\psi}\gamma^{\mu}\psi\overline{\nu_{\alpha}}\gamma_{\mu}P_{L}\nu_{\beta}.\label{eq:L-64}
\end{equation}
Here we adopt the same definition of $y_{\alpha}$ and $y_{\beta}$
as in Eq.~(\ref{eq:L-55}). Similar to the triangle diagram, the result is valid only if $m_{\phi}$ is well above the fermion masses.
 In addition, $y_{\psi}$ is the Yukawa vertex
marked in the box diagram in Fig.~\ref{fig:NSI-loop}. The corresponding
Yukawa interaction is 
\begin{equation}
{\cal L}\supset y_{\psi}\overline{\psi'}\phi\psi+{\rm h.c.},\label{eq:L-68}
\end{equation}
where $\psi$ and $\psi'$ are the external and internal fermion lines
(left part of the box diagram).\\


To summarize, we combine the above loop-induced NSI as
\begin{equation}
{\cal L}_{{\rm NSI}}=\left(\epsilon_{\alpha\beta}^{\triangleright}+\epsilon_{\alpha\beta}^{\boxempty}\right)\frac{G_{F}}{\sqrt{2}}\overline{\psi}\gamma^{\mu}\psi\overline{\nu_{\alpha}}\gamma_{\mu}P_{L}\nu_{\beta},\ \ (\psi=e_{L},\ e_{R},\ u_{L},\ u_{R},\cdots),\label{eq:L-65}
\end{equation}
with the individual contributions 
\begin{equation}
\epsilon_{\alpha\beta}^{\triangleright}=-\frac{8g_{\alpha\beta}^{(1)}}{g}Q_{Z}^{(\psi)}c_{W}, ~~~~
\epsilon_{\alpha\beta}^{\boxempty}=\frac{1}{16\pi^{2}}\frac{\sqrt{2}y_{\alpha}^{*}y_{\beta}|y_{\psi}|^{2}}{4m_{\phi}^{2}G_{F}}.\label{eq:L-67}
\end{equation}
Here $\epsilon_{\alpha\beta}^{\triangleright}$ and $\epsilon_{\alpha\beta}^{\boxempty}$
denote the contributions of the triangle and box diagrams respectively;
$Q_{Z}^{(\psi)}$ is the $Z$ charge of  $\psi$ (electrons/quarks)
as listed in Tab.~\ref{tab:Z-charge}, $g_{\alpha\beta}^{(1)}$ is
given by Eq.~(\ref{eq:L-55}). Note that the fermions considered here are 
chiral. The usually considered vector NSI, cf.\ Eqs.\ (\ref{eq:VALRe}, \ref{eq:VALRnp}) can be obtained by summing for the triangle diagram their corresponding $Q_Z$ charges from Tab.\ \ref{tab:Z-charge}. The box diagram needs to be multiplied by $2$. 
We stress here that since Yukawa couplings can be complex, the various $\epsilon$ can also be complex. 
This is in contrast to typical models in which integrating out a gauge boson generates NSI.



\section{\label{sec:appl}Application to Models}
We will apply the above general results now to  explicit models.

\subsection{Model A:  the minimal charged Higgs model\label{sec:scalar-singlet}}

The first model we consider is a very simple extension of the SM by
adding only a scalar singlet $\phi$ with hypercharge $Y_{\phi}=1$
to the SM. After electroweak symmetry breaking, $\phi$ will eventually
obtain one unit of electric charge. For this reason, we will refer to
the model as the minimal charged Higgs model. The model has been
studied in, e.g., Ref.~\cite{Bilenky:1993bt,Antusch:2008tz,Wise:2014oea} 
(the latter two discuss tree-level NSI effects), and has 
also been considered as a part of larger SM extensions such as the
Zee model \cite{Zee:1980ai}. 

Because the hypercharge is $Y_{\phi}=1$, the only new Yukawa interaction
allowed by symmetry  is $\overline{L^{c}}i\sigma_{2}L\phi$, where
$L=(\nu_{L},\ e_{L})^{T}$ is a SM lepton doublet with hypercharge $Y_{L}=-1/2$;
$L^{c}$ is the charge conjugate of $L$ so $\overline{L^{c}}$ has
the same hypercharge as $L$; $i\sigma_{2}$ is necessary to form
an $SU(2)_{L}$ invariant.  Note that for any two Dirac spinors $\psi_{1}$
and $\psi_{2}$, the combination $\overline{\psi_{1}^{c}}\psi_{2}=\overline{\psi_{2}^{c}}\psi_{1}$
is symmetric  under the interchange of $1\leftrightarrow2$ (similar
to the well-known fact that a Majorana mass matrix is always symmetric).
On the other hand, the $SU(2)_{L}$ product with $i\sigma_{2}$ is
anti-symmetric. As a result, $\overline{L^{c}}i\sigma_{2}L\phi$ vanishes
if the two lepton doublets are of the same flavor: the Yukawa interactions
of $\phi$ can be non-vanishing only when there are at least two different
flavors. Adding the new Yukawa interactions to the SM, the Lagrangian
of this model is 
\begin{align}
{\cal L} & ={\cal L}_{{\rm SM}}+|D_{\mu}\phi|^{2}-m_{\phi}^{2}\phi\phi^{*}-V(\phi,\thinspace H)\label{eq:L-33}\\
 & +\left(\sum_{\alpha,\thinspace\beta}y_{\alpha\beta}\overline{L_{\alpha}^{c}}i\sigma_{2}L_{\beta}\phi+{\rm h.c.}\right),\label{eq:L-34}
\end{align}
where the Yukawa matrix $y_{\alpha\beta}$ is anti-symmetric. The
SM Higgs doublet is denoted as $H$ and $V(\phi,\thinspace H)$ denotes
all quartic terms involving $\phi$ and $H$ together or $\phi$ only.
The scalar mass $m_{\phi}^{2}$ is assumed to be larger than the electroweak
scale to avoid direct constraints from collider searches. For convenience
of later use, we explicitly expand the new Yukawa terms: 
\begin{equation}
\sum_{\alpha,\thinspace\beta}y_{\alpha\beta}\overline{L_{\alpha}^{c}}i\sigma_{2}L_{\beta}\phi=2y_{e\mu}\left(\overline{\nu_{e}^{c}}P_{L}\mu-\overline{\nu_{\mu}^{c}}P_{L}e\right)\phi+2y_{\mu\tau}\left(\overline{\nu_{\mu}^{c}}P_{L}\tau-\overline{\nu_{\tau}^{c}}P_{L}\mu\right)\phi+2y_{\tau e}\left(\overline{\nu_{\tau}^{c}}P_{L}e-\overline{\nu_{e}^{c}}P_{L}\tau\right)\phi+{\rm h.c.}\label{eq:L-42}
\end{equation}
The covariant derivative is
\begin{equation}
D_{\mu}\phi=\partial_{\mu}\phi-ig'B_{\mu}Y_{\phi}\phi,\label{eq:L-36}
\end{equation}
where $B_{\mu}$ is the $U(1)_{Y}$ gauge boson. After the Weinberg
rotation, 
\begin{equation}
\left(\begin{array}{c}
W_{\mu}^{3}\\
B_{\mu}
\end{array}\right)=\left(\begin{array}{cc}
c_{W} & s_{W}\\
-s_{W} & c_{W}
\end{array}\right)\left(\begin{array}{c}
Z_{\mu}\\
A_{\mu}
\end{array}\right),~~
(s_{W},\ c_{W})\equiv\frac{(g',\ g)}{\sqrt{g'^{2}+g^{2}}},\label{eq:L-37}
\end{equation}
we obtain 
\begin{equation}
D_{\mu}\phi=\partial_{\mu}\phi-i\frac{g}{c_{W}}Z_{\mu}Q_{Z}^{(\phi)}\phi-igs_{W}A_{\mu}Q_{A}^{(\phi)}\phi,\label{eq:L-38}
\end{equation}
with the $Z$- and electric charges
\begin{equation}
(Q_{Z}^{(\phi)},\ Q_{A}^{(\phi)})=(-s_{W}^{2},\ 1).\label{eq:L-43}
\end{equation}
Here $Q_{A}^{(\phi)}=1$ implies that $\phi$ has the same electric
charge as the proton, $Q_{Z}^{(\phi)}$ is the $Z$ charge of $\phi$.
The $Z$ charges of the SM fermions have already been defined in Eq.~(\ref{eq:L-56})
and listed in Tab.~\ref{tab:Z-charge}. It is important to notice
that the $Z$ charges in the Yukawa term (\ref{eq:L-42}) are conserved
\begin{equation}
Q_{Z}^{(\phi)}+Q_{Z}^{(\nu_{L})}+Q_{Z}^{(e_{L})}=0,\label{eq:L-41}
\end{equation}
which is crucial for the UV divergences in the relevant loops to cancel. 
Eq.~(\ref{eq:L-41}) is not an accidental result because the
model here is renormalizable and UV divergences should not appear
in any physical processes.

Next, we shall discuss the neutrino NSI in this model, using the general
results which have been obtained in Sec.~\ref{sec:General-analyses}.

\vspace{0.3cm}\noindent {\bf Fierz-transformed NSI:}

\noindent 
We first integrate out $\phi$, which generates the scalar form effective
operator
\begin{equation}
{\cal L}\supset\frac{1}{m_{\phi}^{2}}2y_{\beta e}\left(\overline{\nu_{\beta}^{c}}P_{L}e\right)2y_{\alpha e}^{*}\left(\overline{e}P_{R}\nu_{\alpha}^{c}\right).\label{eq:L-114}
\end{equation}
Because $y_{\alpha\beta}$ is anti-symmetric, $\beta$ and $\alpha$ can only be $\mu$ or $\tau$, but not $e$. So Eq.~(\ref{eq:L-114}) can not generate  NSI between $\nu_e$ and $e$. 
According to Eq.~(\ref{eq:L-51}), with the replacement $(\overline{\psi_{1R}},\ \psi_{2L},\ \overline{\psi_{3L}},\ \psi_{4R})\rightarrow(\overline{\nu_{\beta L}^{c}},\ e_{L},\ \overline{e_{L}},\ \nu_{\alpha L}^{c})$,
we obtain 
\begin{align}
{\cal L} & \supset\frac{4y_{\beta e}y_{\alpha e}^{*}}{m_{\phi}^{2}}\left[-\frac{1}{2}\overline{\nu_{\beta L}^{c}}\gamma^{\mu}\nu_{\alpha L}^{c}\overline{e_{L}}\gamma_{\mu}e_{L}\right]
  =\frac{4y_{\beta e}y_{\alpha e}^{*}}{m_{\phi}^{2}}\left[\frac{1}{2}\overline{\nu_{\alpha L}}\gamma^{\mu}\nu_{\beta_{L}}\overline{e_{L}}\gamma_{\mu}e_{L}\right],\label{eq:L-80}
\end{align}
where in the second step we have used the identity (\ref{eq:L-78}).
Eq.~(\ref{eq:L-80}) is the Fierz-transformed NSI in this model, which is only possible for 
coupling to electrons. We stress the known fact that 
only $\epsilon_{\mu\tau}$, $\epsilon_{\mu\mu}$ and $\epsilon_{\tau\tau}$ can be generated in this model via Fierz transformation, 
and that its magnitude is constrained to be rather small \cite{Antusch:2008tz,Wise:2014oea}, typically around ${\cal O}(10^{-3})$. The strongest constraints are from the variation of $G_F$  extracted from $\mu$ and $\tau$ lifetimes, which are affected because the SM charged current interactions of $\mu$ and $\tau$ are directly modified by the charged Higgs introduced in this model---see Ref.~\cite{Antusch:2008tz} for more detailed analyses.
We will show next that loop-induced NSI terms can generate all flavor terms, though later it turns out that those terms are also constrained to be small. Nevertheless, the analysis illustrates the potential importance of loop effects. \\


\vspace{0.3cm}\noindent {\bf Loop-induced NSI:}

\noindent
Without loss of generality, let us first focus on how $g_{\mu e}^{(1)}$
can be generated according to the results in Sec.~\ref{subsec:Loop-induced-NSI}
and Eq.~(\ref{eq:L-42}). The relevant terms in Eq.~(\ref{eq:L-42})
are 
\begin{equation}
2y_{\mu\tau}\overline{\nu_{\mu}^{c}}P_{L}\tau\phi-2y_{\tau e}\overline{\nu_{e}^{c}}P_{L}\tau\phi=2y_{\mu\tau}\overline{\tau^{c}}P_{L}\nu_{\mu}\phi-2y_{\tau e}\overline{\tau^{c}}P_{L}\nu_{e}\phi.\label{eq:L-82}
\end{equation}
By comparing this expression to Eq.~(\ref{eq:L-53}), we have the mapping
\begin{equation}
\nu_{\alpha}\rightarrow\nu_{\mu},\ \nu_{\beta}\rightarrow\nu_{e},\ \psi_{{\rm int}}\rightarrow\tau^{c};~~
y_{\alpha}^{*}\rightarrow2y_{\mu\tau}^{*},\ y_{\beta}\rightarrow-2y_{\tau e}.\label{eq:L-47}
\end{equation}
Using Eq.~(\ref{eq:L-55}) and assuming $\frac{m_{Z}^{2}}{m_{\phi}^{2}}\ll1$,
we obtain the effective $Z$-$\nu_{e}$-$\nu_{\mu}$ vertex 
\begin{equation}
g_{\mu e}^{(1)}=-\frac{y_{\mu\tau}^{*}y_{\tau e}}{16\pi^{2}}\frac{g}{c_{W}}\frac{m_{Z}^{2}}{m_{\phi}^{2}}\frac{2}{3}\left[\frac{c_{W}^{2}}{3}-(1-2s_{W}^{2})\left(\log\frac{m_{Z}^{2}}{m_{\phi}^{2}}-i\pi\right)\right].\label{eq:L-83}
\end{equation}
For other flavors, one can straightforward derive similar results
accordingly. The general result is 
\begin{equation}
g_{\alpha\beta}^{(1)}=\sum_{\delta=e,\mu,\tau}\frac{y_{\alpha\delta}^{*}y_{\beta\delta}}{16\pi^{2}}\frac{g}{c_{W}}\frac{m_{Z}^{2}}{m_{\phi}^{2}}\frac{2}{3}\left[\frac{c_{W}^{2}}{3}-(1-2s_{W}^{2})\left(\log\frac{m_{Z}^{2}}{m_{\phi}^{2}}-i\pi\right)\right].\label{eq:L-84}
\end{equation}
Eq.~(\ref{eq:L-84}) combined with Eq.~(\ref{eq:L-67}) gives the
triangle NSI in this model:
\begin{equation}
\epsilon_{\alpha\beta}^{\triangleright}=-\frac{8c_{W}}{g}Q_{Z}^{(\psi)}\sum_{\delta}\frac{y_{\alpha\delta}^{*}y_{\beta\delta}}{16\pi^{2}}\frac{g}{c_{W}}\frac{m_{Z}^{2}}{m_{\phi}^{2}}\frac{2}{3}\left[\frac{c_{W}^{2}}{3}-(1-2s_{W}^{2})\left(\log\frac{m_{Z}^{2}}{m_{\phi}^{2}}-i\pi\right)\right].\label{eq:L-103}
\end{equation}
The box NSI in this model also exists, but only for electron-neutrino
NSI because $\psi$ in the right panel of Fig.~\ref{fig:NSI-loop} can only be an electron.
The box NSI parameter $\epsilon_{\alpha\beta}^{\boxempty}$ can be
directly obtained from Eq.~(\ref{eq:L-67}) with the Yukawa couplings
replaced by
\begin{equation}
y_{\alpha}^{*}y_{\beta}\rightarrow\sum_{\delta=e,\mu,\tau}4y_{\alpha\delta}^{*}y_{\beta\delta},\ \ \ |y_{\psi}|^{2}\rightarrow4\left(|y_{e\mu}|^{2}+|y_{e\tau}|^{2}\right),\label{eq:L-85}
\end{equation}
which leads to 
\begin{equation}
\epsilon_{\alpha\beta}^{\boxempty}=\frac{1}{16\pi^{2}}\frac{4\sqrt{2}\sum_{\delta}y_{\alpha\delta}^{*}y_{\beta\delta}}{m_{\phi}^{2}G_{F}}\left(|y_{e\mu}|^{2}+|y_{e\tau}|^{2}\right).\label{eq:L-102}
\end{equation}
Recall that the usually considered vector form for $\epsilon$ is twice the value of Eq.\  (\ref{eq:L-102}). It is noteworthy that all flavor terms $\epsilon_{\alpha \beta}$ can be generated, while the Fierz-transformed NSI was only possible for the $\mu\tau$ case. 

\subsection{Model B: Secret neutrino interactions\label{sec:Model-B}}

Secret neutrino interactions are a type of interactions that only
exist among neutrinos. They are generally difficult to be tested in
terrestrial experiments because electrons and quarks are not involved
in such interactions. However, secret neutrino interactions could
have interesting cosmological and astrophysical effects, in 
supernova dynamics, cosmic neutrino propagation, Big Bang Nucleosynthesis
(BBN), etc. Therefore it has been considered in many references \cite{Kolb:1987qy,Bilenky:1992xn,Hannestad:2013ana,Ng:2014pca,Forastieri:2015paa,Huang:2017egl}.
The simplest secret neutrino interaction is a scalar boson interacting
with the left-handed neutrinos $\phi\nu_{L}\nu_{L}$ where $\nu_{L}$
is in the Weyl spinor notation\footnote{The secret scalar boson could also couple right-handed and left-handed neutrinos together ($\phi\nu_{R}\nu_{L}$), which has different phenomenological consequences in cosmological and astrophysical processes. In this case, due to the absence of $Z$ coupling to $\nu_R$, there is no loop-induced NSI.}. In the Dirac notation, and including
the flavor indices, the interaction should be formulated as 
\begin{equation}
{\cal L}\supset y_{\alpha\beta}\phi\overline{\nu_{\alpha L}^{c}}\nu_{\beta L}+{\rm h.c.}\label{eq:L-86}
\end{equation}
We demonstrate now that the secret neutrino interaction in
Eq.~(\ref{eq:L-86}) leads to loop-induced NSI. No NSI are generated when the scalar is integrated out. 
Because $\phi$
does not couple to charged fermions in this model, the Fierz-transformed
NSI and the loop-induced NSI from the box diagram are absent. Only
the triangle diagram can generate NSI. 

By comparing Eq.~(\ref{eq:L-86}) to Eq.~(\ref{eq:L-53}), we can
use the mapping 
\begin{equation}
y_{\alpha}^{*}y_{\beta}\rightarrow\sum_{\delta}y_{\delta\alpha}^{*}y_{\delta\beta},\ \psi_{{\rm int}}\rightarrow\nu_{\alpha L}^{c},\ Q_{Z}^{(\psi_{{\rm int}})}\rightarrow-Q_{Z}^{(\nu_{L})},\label{eq:L-88}
\end{equation}
to find {[}cf.\ Eq.~(\ref{eq:L-55}){]}:
\[
g_{\alpha\beta}^{(1)}=\frac{1}{16\pi^{2}}\frac{g}{c_{W}}Q_{Z}^{(\nu_{L})}\frac{m_{Z}^{2}}{m_{\phi}^{2}}\sum_{\delta}y_{\delta\alpha}^{*}y_{\delta\beta}\left[f(r)-h(r)\right].
\]
Then the corresponding triangle NSI parameter in Eq.~(\ref{eq:L-67}) is:
\begin{equation}
\epsilon_{\alpha\beta}^{\triangleright}=-\frac{4}{16\pi^{2}}Q_{Z}^{(\psi)}\frac{m_{Z}^{2}}{m_{\phi}^{2}}\sum_{\delta}y_{\delta\alpha}^{*}y_{\delta\beta}\left[f(r)-h(r)\right].\label{eq:L-87}
\end{equation}
Note that the internal fermions in the triangle diagram are left-handed neutrinos, and recall that 
$r = m_Z^2/m_\phi^2$. 
However, one should note that Eq.~(\ref{eq:L-86}) is not a complete
model so the UV divergences cannot be fully canceled without introducing
new particles or new interaction terms. 
Consequently there is a UV divergence, explicitly
shown in Eq.~(\ref{eq:L-15}) and not given here. In a complete and renormalizable model
containing the secret neutrino interaction (\ref{eq:L-86}), this
UV divergence will be canceled by additional diagrams, potentially modifying 
 the result (\ref{eq:L-87}). 
Since this depends on the details of the complete model, we refrain from 
going further into detail and keep Eq.~(\ref{eq:L-87}), which should 
be order-of-magnitude wise correct. 
Regarding Eq.~(\ref{eq:L-86}) there is not necessarily lepton number violation because $\phi$ could carry two units of lepton number. However, if the lepton number is violated by, e.g., non-zero $\left\langle \phi\right\rangle$, then such a secret interaction can also be responsible for a Majorana neutrino mass.  If this term is the only term responsible for neutrino mass, it is interesting to note that 
$\epsilon_{\alpha\beta}^{\triangleright} \propto (m_\nu m_\nu^\dagger)_{\alpha \beta}$, which would result in $\epsilon_{e\mu}^{\triangleright}  \simeq \epsilon_{e\tau}^{\triangleright} \ll 
\epsilon_{\mu\tau}^{\triangleright} $, where the proportionality factor between  
$\epsilon_{e\mu}^{\triangleright}$ and $\epsilon_{\mu\tau}^{\triangleright} $ is about 
$\Delta m^2_{21}/|\Delta m^2_{31}|$ \cite{Chakrabortty:2012vp}. 

It should be noticed that in the UV complete models containing the secret neutrino interactions, $\phi$ may or may not be accompanied with a charged Higgs, depending on whether $\phi$ is the neutral component of an $SU(2)_L$ multiplet or not. The former case usually suffers from stringent constraints due to its connection with the charged Higgs---see, e.g., \cite{Dey:2018yht}. In the latter case, the secret neutrino interactions are usually obtained by mass mixing of left-handed neutrinos with other singlet fermions such as right-handed neutrinos, which happens in the Majoron model \cite{Chikashige:1980ui} and its variants \cite{Valle:1990pka}. For such models, one needs to check whether the sizable mixing would lead to correct light neutrino masses or not.
Since all these details are very model-dependent, we refrain here from further discussions on the UV complete models of secret neutrino interactions.

Due to a lack of very stringent terrestrial constraints on the secret neutrino interactions,
the loop-induced NSI in this model can be in principle much larger
than in the previous model. We will discuss possible sizes of the NSI later in Sec.\ \ref{sec:How-large}. 

\subsection{Model C: Neutral scalar boson\label{sec:Model-C}}

Neutrinos could also have new scalar interactions with the charged
fermions mediated by a neutral scalar, which can be expressed 
by the following Lagrangian: 
\begin{equation}
{\cal L}\supset y_{\alpha\beta}^{\nu}\phi\overline{\nu_{\alpha}}\nu_{\beta}+y^{\psi}\phi\overline{\psi}\psi+{\rm h.c.}\ \ (\psi=e,\ u,\ d).\label{eq:L-89}
\end{equation}
Since neutrino-electron and coherent elastic neutrino-nucleus scattering are 
induced, Eq.~(\ref{eq:L-89}) has interesting phenomenological impact 
 on experiments such 
as COHERENT \cite{Akimov:2017ade}, CONUS \cite{CONUS}, 
CHARM II \cite{Vilain:1993kd,Vilain:1994qy}, LSND \cite{Auerbach:2001wg}, TEXONO \cite{Deniz:2009mu},  GEMMA \cite{Beda:2009kx,Beda:2010hk}, etc., 
see e.g.\ Refs.\ \cite{Lindner:2016wff,Lindner:2018kjo,Farzan:2018gtr}. 

In this model, because the scalar boson is neutral, there is no Fierz-transformed
NSI. In the triangle diagram (Fig.~\ref{fig:NSI-loop}), since the
external neutrino lines are left-handed neutrinos, the internal fermion
lines can only be right-handed neutrinos\footnote{One may also consider another type of $\phi$-$\nu$ interaction similar to Eq.~(\ref{eq:L-86}). In this case, the loop-induced NSI is a combination of model B and model C---it has the same $\epsilon_{\alpha\beta}^{\triangleright}$ as model B and the same $\epsilon_{\alpha\beta}^{\boxempty}$  as model C.
} because $\phi\overline{\nu_{\alpha}}\nu_{\beta}=\phi\overline{\nu_{\alpha R}}\nu_{\beta L}+\phi\overline{\nu_{\alpha L}}\nu_{\beta R}$.
Since right-handed neutrinos do not couple to the $Z$ boson 
there is no triangle NSI in this model, i.e. 
\begin{equation}
\epsilon_{\alpha\beta}^{\triangleright}=0.\label{eq:L-90}
\end{equation}
However, since left- and right-handed elecrons and quarks do exist, this model leads to triangle diagrams correcting their couplings to the $Z$ boson. Therefore, the model is constrained by the partial decay widths of $Z$. We will discuss this issue in Sec.~\ref{sec:How-large}.

On the other hand, this model has loop-induced NSI from the box diagram. By
comparing Eq.~(\ref{eq:L-89}) to Eq.~(\ref{eq:L-53}) and (\ref{eq:L-68}),
we have the mapping 
\begin{equation}
y_{\alpha}^{*}y_{\beta}\rightarrow\sum_{\delta=e,\mu,\tau}y_{\alpha\delta}^{\nu}y_{\delta\beta}^{\nu},\ \ |y_{\psi}|^{2}\rightarrow|y^{\psi}|^{2},\label{eq:L-91}
\end{equation}
which gives the box NSI parameter:
\begin{equation}
\epsilon_{\alpha\beta}^{\boxempty}=\frac{1}{16\pi^{2}}\frac{\sqrt{2}|y^{\psi}|^{2}}{4m_{\phi}^{2}G_{F}}\sum_{\delta=e,\mu,\tau}y_{\alpha\delta}^{\nu}y_{\delta\beta}^{\nu}.\label{eq:L-67-1}
\end{equation}
Although Eq.~(\ref{eq:L-89}) is not a complete model, in contrast
to model B, there is no UV divergence in computing the loop-induced
NSI because the box diagram is always finite.

\section{How large can loop-induced NSI be?\label{sec:How-large}}

Now that we have derived loop-induced NSI both in the general framework
and in several specific models, a natural question to ask is how large
they can be. The answer of course depends on the models as well as
the experimental constraints. In this section, we summarize some experimental
constraints on the three models and estimate the allowed magnitude 
of loop-induced vector NSI for couplings to electrons, protons and neutrons, whose definition is given in Eqs.\ (\ref{eq:VALRe}, \ref{eq:VALRnp}). 
We selectively consider three most relevant experimental
constraints, namely the invisible $Z$ decay width, elastic neutrino
scattering and charged lepton flavor violation. When all these constraints
are taken into consideration, we find that loop-induced NSI in the
three models can reach the magnitude listed in Tab.~\ref{tab:NSI}.

\begin{table*}
\caption{\label{tab:NSI}Reachable magnitude of the Fierz and loop-induced NSI in the
three models under study. Here $\epsilon^{F}$, $\epsilon^{\triangleright}$ and  $\epsilon^{\boxempty}$ are generated by Fierz transformations, triangle and box diagrams respectively.}
\begin{ruledtabular}
\begin{minipage}{0.9\textwidth}

\begin{tabular}{cccccccc}
 & $\epsilon_{e}^{F}$ & $\epsilon_{e}^{\triangleright}$ & $\epsilon_{n}^{\triangleright}$ & $\epsilon_{p}^{\triangleright}$ & $\epsilon_{e}^{\boxempty}$ & $\epsilon_{n}^{\boxempty}$ & $\epsilon_{p}^{\boxempty}$\tabularnewline
\hline 
model A & ${\cal O}(10^{-3})$ & ${\cal O}(10^{-5})$ & ${\cal O}(10^{-4})$ & ${\cal O}(10^{-5})$ & ${\cal O}(10^{-3})$ & 0 & 0\tabularnewline
model B & 0 & ${\cal O}(10^{-1})$ & ${\cal O}(1)$ & ${\cal O}(10^{-1})$ & 0 & 0 & 0\tabularnewline
model C & 0 & 0 & 0 & 0 & ${\cal O}(10^{-2})$ & ${\cal O}(10^{-2})$ & ${\cal O}(10^{-2})$\tabularnewline
\end{tabular}

\end{minipage}
\end{ruledtabular}

\end{table*}


\vspace{0.3cm}\noindent {\bf  Invisible $Z$ decay width}


\noindent
Since in the triangle diagram the $Z\overline{\nu}\nu$ vertices are
modified in models A and B, it is necessary to consider the effect on the invisible
$Z$ decay width which has been measured precisely \cite{ALEPH:2005ab}:
\begin{equation}
\Gamma_{Z, {\rm inv}}=N_{\nu}\Gamma_{Z\rightarrow\nu\overline{\nu}},\ \ N_{\nu}=2.9840\pm0.0082.\label{eq:L-92}
\end{equation}
Adding Eq.~(\ref{eq:L-55}) to the SM $Z\overline{\nu}\nu$ terms,
we have the following $Z\overline{\nu}\nu$ interactions
\begin{equation}
{\cal L}_{Z\overline{\nu}\nu}=\frac{gQ_{Z}^{(\nu_{L})}}{c_{W}}Z_{\mu}\lambda_{\alpha\beta}\overline{\nu_{\alpha}}\gamma^{\mu}P_{L}\nu_{\beta},
\mbox{ with } 
\lambda_{\alpha\beta}=\frac{g_{\alpha\beta}^{(1)}}{g}\frac{c_{W}}{Q_{Z}^{(\nu_{L})}}+\delta_{\alpha\beta}.\label{eq:L-94}
\end{equation}
For one generation of neutrinos, the decay width $\Gamma_{Z\rightarrow\nu\overline{\nu}}$
is proportional to the absolute square of the vertex coupling. Generalizing
to three generations, it holds that  $\Gamma_{Z, {\rm inv}}\propto{\rm tr}[\lambda\lambda^{\dagger}]$, from which we can infer 
\begin{equation}
N_{\nu}={\rm tr}[\lambda\lambda^{\dagger}]=\sum_{\alpha,\beta}|\lambda_{\alpha\beta}|^{2}.\label{eq:L-95}
\end{equation}
Therefore, the invisible $Z$ decay width should give a strong constraint
on ${\rm tr}[\lambda\lambda^{\dagger}]$. However, one should note
that even when ${\rm tr}[\lambda\lambda^{\dagger}]$ is fixed at $3$,
large values of $g_{\alpha\beta}^{(1)}$ are still allowed due to cancellations in the matrix product. 
The constraint from invisible $Z$ decay is only useful
when it is combined with the elastic neutrino scattering constraints
to be introduced next.

\vspace{0.3cm}\noindent {\bf  Elastic neutrino scattering}

\noindent
New neutrino interactions can be directly constrained by elastic neutrino
scattering experiments \cite{Deniz:2017zok,Farzan:2018gtr,Lindner:2016wff,Lindner:2018kjo}.
Some neutrino-electron scattering experiments (e.g.\ CHARM II \cite{Vilain:1993kd,Vilain:1994qy},
LSND \cite{Auerbach:2001wg}, TEXONO \cite{Deniz:2009mu}) already have precision measurement of
the SM process and most recently coherent elastic neutrino-nucleus
scattering has been successfully observed and will also be precisely
measured in the near future \cite{Akimov:2017ade,CONUS}. 

In general when there are new neutrino interactions, elastic neutrino
scattering is sensitive to the ratio between the new and SM cross
sections (ignoring spectral effects):
\begin{equation}
R_{\alpha}\equiv\frac{\sigma_{{\rm new}}(\nu_{\alpha}+\psi\rightarrow\nu+\psi)}{\sigma_{{\rm SM}}(\nu_{\alpha}+\psi\rightarrow\nu_{\alpha}+\psi)},\label{eq:L-104}
\end{equation}
where  the final neutrino state in the numerator can be of any flavor 
and the cross section $\sigma_{{\rm new}}$ is a sum over all possible
flavors. The target particle $\psi$ can be either an electron or
a nucleus. 

Considering the specific models in this paper, the $Z\overline{\nu}\nu$
vertices are modified in model A and model B, while in model A and
model C the scalar bosons make tree-level contributions to neutrino-electron/nucleus
scattering. 

Given the modified $Z\overline{\nu}\nu$ vertices in Eq.~(\ref{eq:L-94}),
 it is straightforward to derive\footnote{For $\nu_{e}+e$ scattering, there are also $W^{\pm}$ (charged current)
contributions, which can be taken into account by replacing $Q_{Z}^{(\nu_{L})}$
in Eq.~(\ref{eq:L-94}) with an effective value. For simplicity, in
this paper we do not consider this part of contributions in our 
estimation of experimental constraints.}
\begin{equation}
R_{\alpha}=\sum_{\beta}|\lambda_{\alpha\beta}|^{2}.\label{eq:L-96}
\end{equation}
It is interesting to note that Eq.~(\ref{eq:L-95}) can be expressed
in terms of the ratios $R_{\alpha}$: 
\begin{equation}
N_{\nu}={\rm tr}[\lambda\lambda^{\dagger}]=R_{e}+R_{\mu}+R_{\tau}.\label{eq:L-107}
\end{equation}
According to the $\nu_{\mu}$ and $\nu_{e}$ elastic scattering data
\cite{Vilain:1993kd,Auerbach:2001wg,Deniz:2009mu}, $R_{e}$ and $R_{\mu}$ cannot have large
deviations from $1$: 
\begin{equation}
\delta R_{e}\equiv|R_{e}-1|\lesssim20\%,\ \ \delta R_{\mu}\equiv|R_{\mu}-1|\lesssim3\%.\label{eq:L-108}
\end{equation}
This combined with the $Z$ decay observation $|N_{\nu}-3|\ll1$ implies
that $R_{\tau}$ should also be close to $1$:
\begin{equation}
\delta R_{\tau}\equiv|R_{\tau}-1|\lesssim20\%.\label{eq:L-109}
\end{equation}
Using Eq.~(\ref{eq:L-94}), we can convert\footnote{More explicitly, we first substitute the expression of $\lambda_{\alpha\beta}$ in Eq.~(\ref{eq:L-94}) into Eq.~(\ref{eq:L-96}) and then 
check the maximally allowed value of each $|\lambda_{\alpha\beta}-\delta_{\alpha\beta}|$ individually with $R_{\alpha}$ varying in $[1-\delta R_{\alpha},\ 1+\delta R_{\alpha}]$.}
 the constraints on $R_{\alpha}$
to constraints on $g_{\alpha\beta}^{(1)}$: 
\begin{equation}
\left|\frac{g_{\alpha\beta}^{(1)}}{g}\frac{c_{W}}{Q_{Z}^{(\nu_{L})}}\right|<\delta_{\alpha\beta}+\sqrt{1+\delta R_{\alpha}}.\label{eq:L-106}
\end{equation}
Thus, using Eq.~(\ref{eq:L-67}), the corresponding  constraints on $\epsilon_{\alpha\beta}^{\triangleright}$
are 
\begin{equation}
|\epsilon_{\alpha\beta}^{\triangleright}| < 
4(\delta_{\alpha\beta}+\sqrt{1+\delta R_{\alpha}})|Q_{Z}^{(\psi)}|=\begin{cases}
{\cal O}(0.1) & (\psi=e\ {\rm {\rm or}}\ p)\\
{\cal O}(1) & (\psi=n)
\end{cases},\label{eq:L-110}
\end{equation}
where for $\psi=e$ or $p$ the result is suppressed by their small
$Z$ charges $|Q_{Z}^{(e)}|=|Q_{Z}^{(p)}|\propto1-4s_{W}^{2}$. To obtain the ${\cal O}(1)$ NSI in model B referred to in Tab.~\ref{tab:NSI}, we can take, for example, $m_{\phi}=100\ {\rm MeV}$ and $y=10^{-2}$, which according to Eq.~(\ref{eq:L-87}) should lead to $\epsilon_{\alpha\beta}^{\triangleright}={\cal O}(1)$. Model A, as we discuss below, cannot induce such large NSI due to further constraints from charged lepton flavor violation.

The tree-level contribution of the scalar boson in model C is roughly
\begin{equation}
\delta R_{\alpha}\simeq\sum_{\beta}\left(\frac{y_{\alpha\beta}^{\nu}y^{\psi}}{m_{\phi}^{2}}\right)^{2}/\left(\frac{g^{2}}{m_{Z}^{2}c_{W}^{2}}\right)^{2}.\label{eq:L-111}
\end{equation}
Assuming $y_{\alpha\beta}^{\nu}y^{\psi}$ is ${\cal O}(1)$,
the box NSI in model C could reach
\begin{equation}
\epsilon_{\alpha\beta}^{\boxempty}\simeq\frac{1}{8\pi^{2}}\sqrt{3\delta R_{\alpha}}\simeq\begin{cases}
1.0\times10^{-2} & ({\rm if\ }\delta R_{\alpha}=20\%)\\
3.8\times10^{-3} & ({\rm if\ }\delta R_{\alpha}=3\%)
\end{cases}.\label{eq:L-105}
\end{equation}
We mentioned before that model C is also constrained by the partial decay widths of $Z$ to $e$, $u$, and $d$. Here we can check that the constraint in Eq.~(\ref{eq:L-105}) is consistent with current uncertainties of the partial decay widths.
Still assuming Yukawa couplings to be one, with Eq.~(\ref{eq:L-67-1}), we obtain
\begin{equation}
\frac{3\sqrt{2}}{8m_\phi^2 G_F} = 3\frac{c_W^2}{g^2}\frac{m_Z^2}{m_\phi^2}\simeq \sqrt{3\delta R_\alpha}.
\end{equation}
This implies
\begin{equation}
\frac{m_Z^2}{m_\phi^2}\simeq 0.14\,,\quad m_{\phi}^2\simeq 5.9\times10^4\  {\rm GeV}, 
\end{equation}
which can be used to study the corrected $Z$ couplings of $\psi=e,u,d$ from triangle diagrams. We slightly modify Eq.~(\ref{eq:L-55}) to describe the correction to the left-handed coupling $Q_Z^{\psi_L}$ to $Z$ by
\begin{equation}
\label{eq:modZcouplings}
{\cal L}_{{\rm eff}}=g_{\psi_L}^{(1)}Z_{\mu}\overline{\psi}\gamma^{\mu}P_{L}\psi,\ 
\mbox{ with }\ 
g_{\psi_L}^{(1)}=
\frac{y_{\psi}^{*}y_{\psi}}{16\pi^{2}}\frac{g^3}{3c_{W}^3}\sqrt{3\delta R_{\alpha}}\left[f(r)Q_{Z}^{(\psi_{L})}+h(r)Q_{Z}^{(\psi_R)}\right],
\end{equation}
and likewise for the right-handed coupling.
Estimating orders of magnitude, 
\begin{equation}
1/48\pi^2\approx3\times10^{-3},\quad g/c_W\approx1,\quad \sqrt{3\delta R_{\alpha}}\approx 10^{-1},\quad 
|f(r)| \approx 8\times 10^{-3}, \quad |h(r)| \approx 0.2,
\end{equation}
we get an order $10^{-5}$ correction to the coupling in the case of the $h(r)$ term. We conclude that the SM couplings to $Z$ get corrected at the order of $10^{-5}$, which would also be the order of corrections to decay amplitudes. The partial decay widths of $Z$ are known to about 3-digit accuracy \cite{Tanabashi:2018oca}, such that the bounds from neutrino-electron scattering are slightly more stringent.

Similar constraints also exist for model A, which should be approximately
of the same magnitude. However, as we will see, the constraints from
charged lepton decay are much more stringent than those from elastic
neutrino scattering in model A.

\vspace{0.3cm}\noindent {\bf  Charged lepton flavor violation}

\noindent
Charged lepton flavor violation (CLFV) could cause rare lepton decays
such as $\mu\rightarrow e\gamma$, $\mu\rightarrow3e$, $\tau\rightarrow\mu\gamma$,
etc. Currently all lepton flavor violating decays have not been observed,
which yields very strong constraints on models containing CLFV. 
In this paper, we only need to consider CLFV in model A because the
other two models only have flavor violations limited to the neutrino
sector. Here we would like to refer to Ref.~\cite{Antusch:2008tz}
which has studied these decay processes in model A. We present
the results in Ref.~\cite{Antusch:2008tz} with the experimental
bounds updated. 

The CLFV decay widths in model A are given by 
\begin{equation}
\Gamma(\ell_{\alpha}\rightarrow\ell_{\beta}\gamma)=\frac{1}{16\pi^{2}}\frac{g^{2}s_{W}^{2}}{12}\left|\frac{\sum_{\delta}y_{\alpha\delta}y_{\beta\delta}^{*}}{m_{\phi}^{2}G_{F}}\right|^{2}\Gamma(\ell_{\alpha}\rightarrow\nu_{\alpha}\ell_{\beta}\overline{\nu_{\beta}}),\label{eq:L-97}
\end{equation}
\begin{equation}
\Gamma(\ell_{\alpha}\rightarrow\ell_{\beta}\ell_{\beta'}\overline{\ell_{\beta'}})=\frac{c_{W}^{2}}{g^{2}}\left|2g_{\alpha\beta}^{(1)}Q_{Z}^{(e_{L})}\right|^{2}\Gamma(\ell_{\alpha}\rightarrow\nu_{\alpha}\ell_{\beta}\overline{\nu_{\beta}}),\label{eq:L-98}
\end{equation}
where $\ell_{\alpha}\rightarrow\nu_{\alpha}\ell_{\beta}\overline{\nu_{\beta}}$
is a SM charged current process. For example, the following branching 
ratios  have been precisely measured \cite{Patrignani:2016xqp}:
\begin{equation}
{\rm Br}(\mu\rightarrow\nu_{\mu}e\overline{\nu_{e}})\approx100\%,\ {\rm Br}(\tau\rightarrow\nu_{\tau}e\overline{\nu_{e}})=(17.82\pm0.04)\%,\ {\rm Br}(\tau\rightarrow\nu_{\tau}\mu\overline{\nu_{\mu}})=(17.39\pm0.04)\%.\label{eq:L-101}
\end{equation}
The branching ratios with CLFV are highly suppressed, 
the following limits at 90\% CL exist \cite{Patrignani:2016xqp}:
\begin{align}
 & {\rm Br}(\mu\rightarrow e\gamma)<5.7\text{\texttimes}10^{-13},\ \ {\rm Br}(\tau\rightarrow e\gamma)<3.3\text{\texttimes}10^{-8},\ \ {\rm Br}(\tau\rightarrow\mu\gamma)<4.4\text{\texttimes}10^{-8},\label{eq:L-100}\\
 & {\rm Br}(\mu\rightarrow3e)<1.0\text{\texttimes}10^{-12},\ \ {\rm Br}(\tau\rightarrow3e)<2.7\text{\texttimes}10^{-8},\ \ {\rm Br}(\tau\rightarrow3\mu)<2.1\text{\texttimes}10^{-8}.\label{eq:L-99}
\end{align}
From the above data, we can derive the corresponding constraints on
$g_{\alpha\beta}^{(1)}$ according to Eqs.~(\ref{eq:L-97}), (\ref{eq:L-98}),
and (\ref{eq:L-84}):
\begin{align}
 & |g_{\mu e}^{(1)}|<8.9\text{\texttimes}10^{-8}\ (\mu\rightarrow e\gamma),\ \ |g_{\tau e}^{(1)}|<5.0\text{\texttimes}10^{-5}\ (\tau\rightarrow e\gamma),\ \ |g_{\tau\mu}^{(1)}|<6.0\text{\texttimes}10^{-5}\ (\tau\rightarrow\mu\gamma),\label{eq:L-100-1}\\
 & |g_{\mu e}^{(1)}|<1.3\text{\texttimes}10^{-6},\ (\mu\rightarrow3e),\ \ |g_{\tau e}^{(1)}|<5.1\text{\texttimes}10^{-4}\ (\tau\rightarrow3e),\ \ |g_{\tau\mu}^{(1)}|<4.5\text{\texttimes}10^{-4}\ (\tau\rightarrow3\mu).\label{eq:L-99-1}
\end{align}
Note that those bounds have a weak dependence
on $m_{\phi}^{2}$ due to the $\log\frac{m_{Z}^{2}}{m_{\phi}^{2}}$
term in Eq.~(\ref{eq:L-84}). For simplicity, we have set $m_{\phi}=500\ {\rm GeV}$. 
 As one can see,
the constraints in Eq.~(\ref{eq:L-99-1}) are weaker than Eq.~(\ref{eq:L-100-1}).
Taking the values in Eq.~(\ref{eq:L-100-1}), we get 
\begin{equation}
|\epsilon_{\mu e}^{\triangleright}|<1.0\times10^{-6} \, Q_{Z}^{(\psi)},\ \ |\epsilon_{\tau e}^{\triangleright}|<5.5\times10^{-4} \, Q_{Z}^{(\psi)},\ \ |\epsilon_{\tau\mu}^{\triangleright}|<6.5\times10^{-4}\, Q_{Z}^{(\psi)},\label{eq:L-112}
\end{equation}
where $Q_{Z}^{(\psi)}=s_{W}^{2}-\frac{1}{4}$, $-\frac{1}{4}$, or
$\frac{1}{4}-s_{W}^{2}$, for $\psi=e$, or $n$, or $p$ respectively.

Similar to $\epsilon_{\mu e}^{\triangleright}$, the CLFV constraints
on $\epsilon_{\alpha\beta}^{\boxempty}$ from $\ell_{\alpha}\rightarrow\ell_{\beta}\gamma$
are also more stringent than those from $\ell_{\alpha}\rightarrow\ell_{\beta}\ell_{\beta'}\overline{\ell_{\beta'}}$.
According to Eq.~(\ref{eq:L-97}) and Eq.~(\ref{eq:L-102}), the bounds
in Eq.~(\ref{eq:L-100}) cannot be directly converted to the bounds
on $\epsilon_{\alpha\beta}^{\boxempty}$ without known bounds on $|y_{e\mu}|^{2}+|y_{e\tau}|^{2}$.
So for simplicity, we set $|y_{e\mu}|^{2},\ |y_{e\tau}|^{2}<1$, and
get 
\begin{equation}
\epsilon_{\mu e}^{\boxempty}<7.8\times10^{-6},\ \ \epsilon_{\tau e}^{\boxempty}<4.4\times10^{-3},\ \ \epsilon_{\tau\mu}^{\boxempty}<5.2\times10^{-3}.\label{eq:L-113}
\end{equation}

Combining all the constraints discussed above, the strongest constraints
on the loop-induced NSI parameters come from CLFV for model A, and
elastic neutrino scattering for modela B and C. The results are summarized
in Tab.~\ref{tab:NSI}. 


\section{Conclusion\label{sec:Conclusion}}

In scalar extensions of the SM, complex NSI can be generated at the loop level,
denoted here loop-induced NSI. There are two types of loop diagrams that
are responsible for loop-induced NSI, triangle diagrams and 
box diagrams shown in Fig.~\ref{fig:NSI-loop}. We computed the loop
diagrams and derived general formulae for loop-induced NSI, given
by Eqs.~(\ref{eq:L-55}) and (\ref{eq:L-67}). 

To be more concrete, we applied our results to three specific and frequently 
discussed models, which contain charged or neutral scalar bosons. 
With the experimental constraints 
on these models taken into consideration, we estimated how large the
loop-induced NSI can be, which is summarized in
Tab.~\ref{tab:NSI}. Testable NSI are possible. 

Our calculations were performed in the limit of heavy scalars (heavier than the fermion masses in loops), though a similar analysis could also be performed for light particles.  
Loop-induced NSI are not neccesarily obtainable by scalar particles  only, but 
also by vector bosons, leptoquarks etc. The relevant phenomenology will differ and 
deserves future study. 

\begin{acknowledgments}
IB is supported by the IMPRS-PTFS and enrolled at 
Heidelberg University. WR is supported by the DFG with grant RO 2516/6-1 in 
the Heisenberg program.
\end{acknowledgments}

\appendix

\section{The triangle diagrams \label{sec:triangle-diagrams}}

In this Appendix, we compute the triangle diagrams in a general $U(1)$
model. The result can be directly applied to more complicated models
such as the SM extended by various scalar particles. Various useful identities and relations necessary for our calculations can be found in Appendix \ref{sec:appC}.

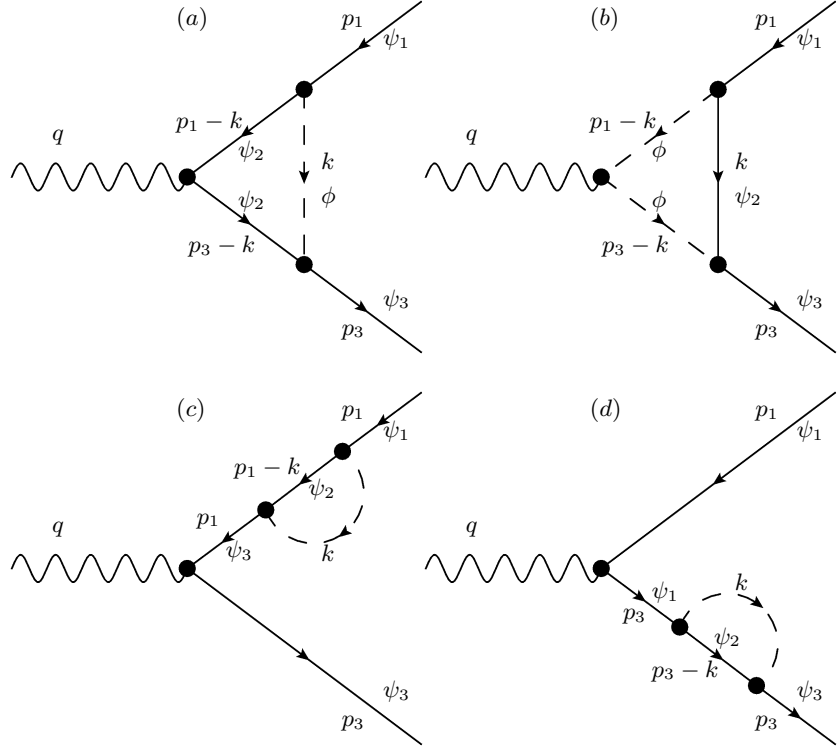
\begin{figure}[t]
\centering

\begin{overpic}[width=5.5cm]
{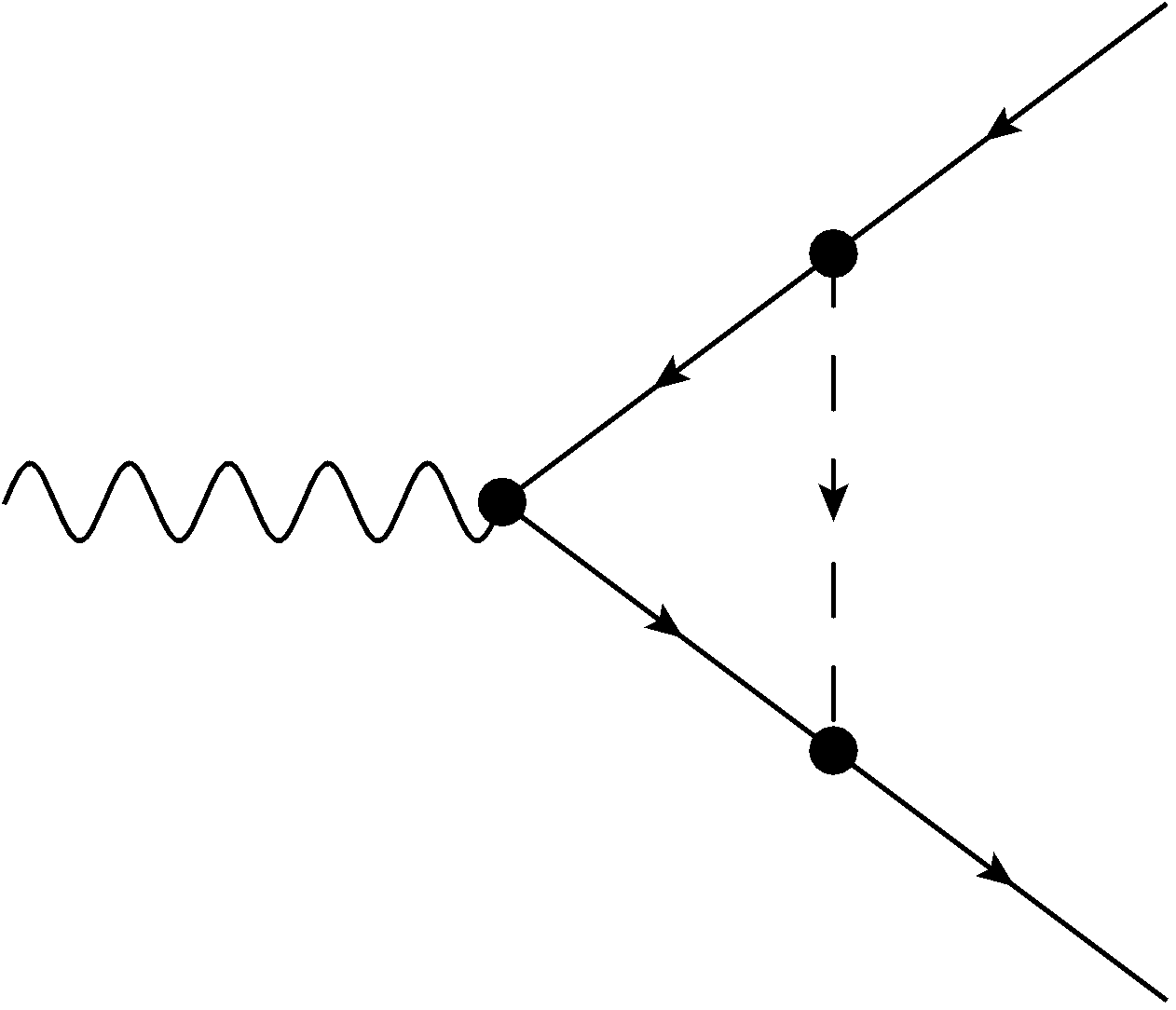} 
\put (40,80) {$(a)$}
\put (10,52) {$q$} 
\put (80,80) {$p_{1}$} \put (80,5) {$p_{3}$}
\put (75,45) {$k$}

\put (40,55) {$p_1-k$}
\put (43,25) {$p_3-k$}

\put (90,75) {$\psi_1$} \put (90,12) {$\psi_3$} 
\put (55,48) {$\psi_2$} \put (55,36) {$\psi_2$} 
\put (75,37) {$\phi$}

\end{overpic}
\begin{overpic}[width=5.5cm]
{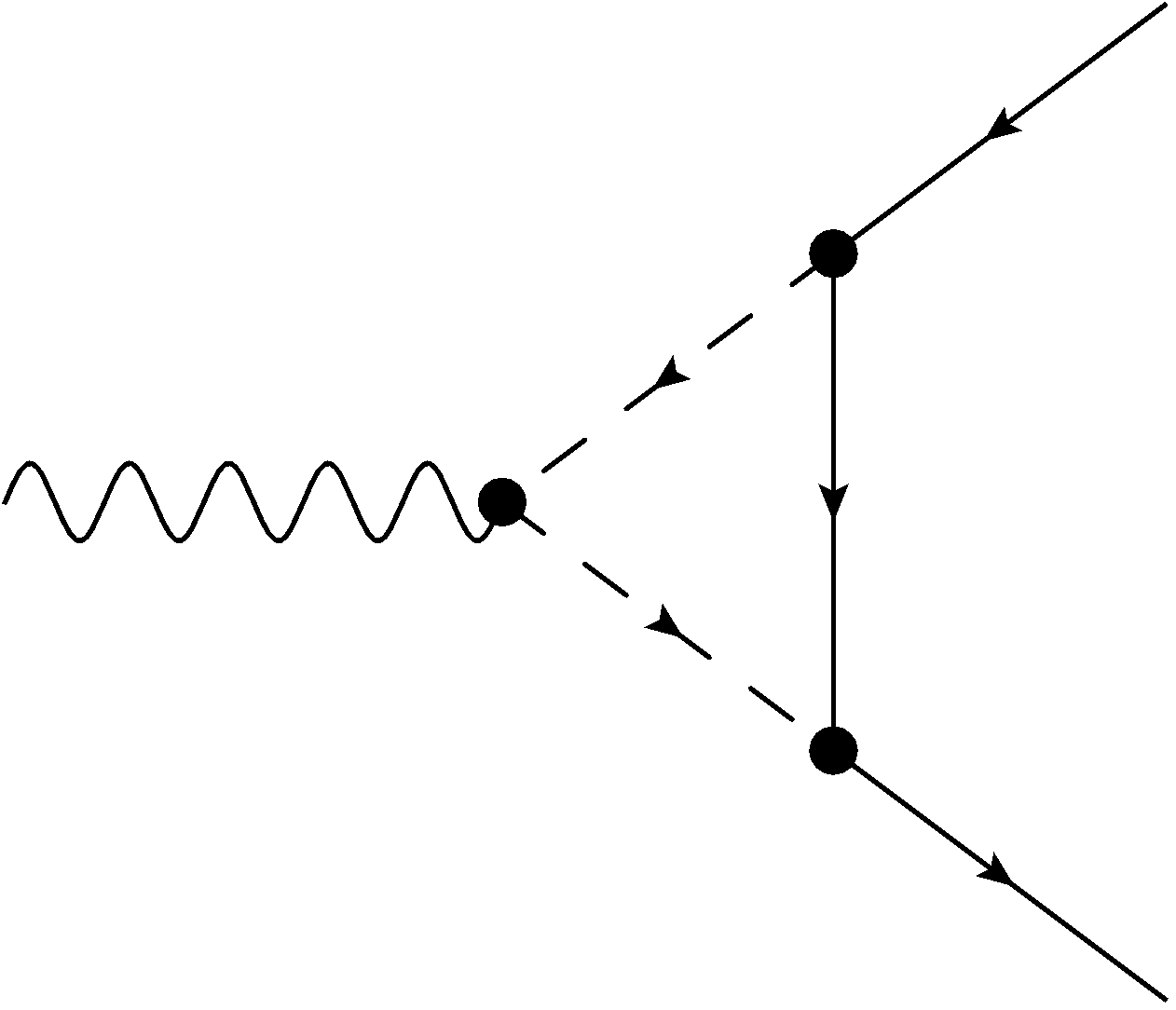} 
\put (40,80) {$(b)$}
\put (10,52) {$q$} 
\put (80,80) {$p_{1}$} \put (80,5) {$p_{3}$}
\put (75,45) {$k$}

\put (40,55) {$p_1-k$}
\put (43,25) {$p_3-k$}

\put (90,75) {$\psi_1$} \put (90,12) {$\psi_3$} 
\put (55,48) {$\phi$} \put (55,36) {$\phi$} 
\put (75,37) {$\psi_2$}

\end{overpic}

\vspace{0.5cm}

\begin{overpic}[width=5.5cm]
{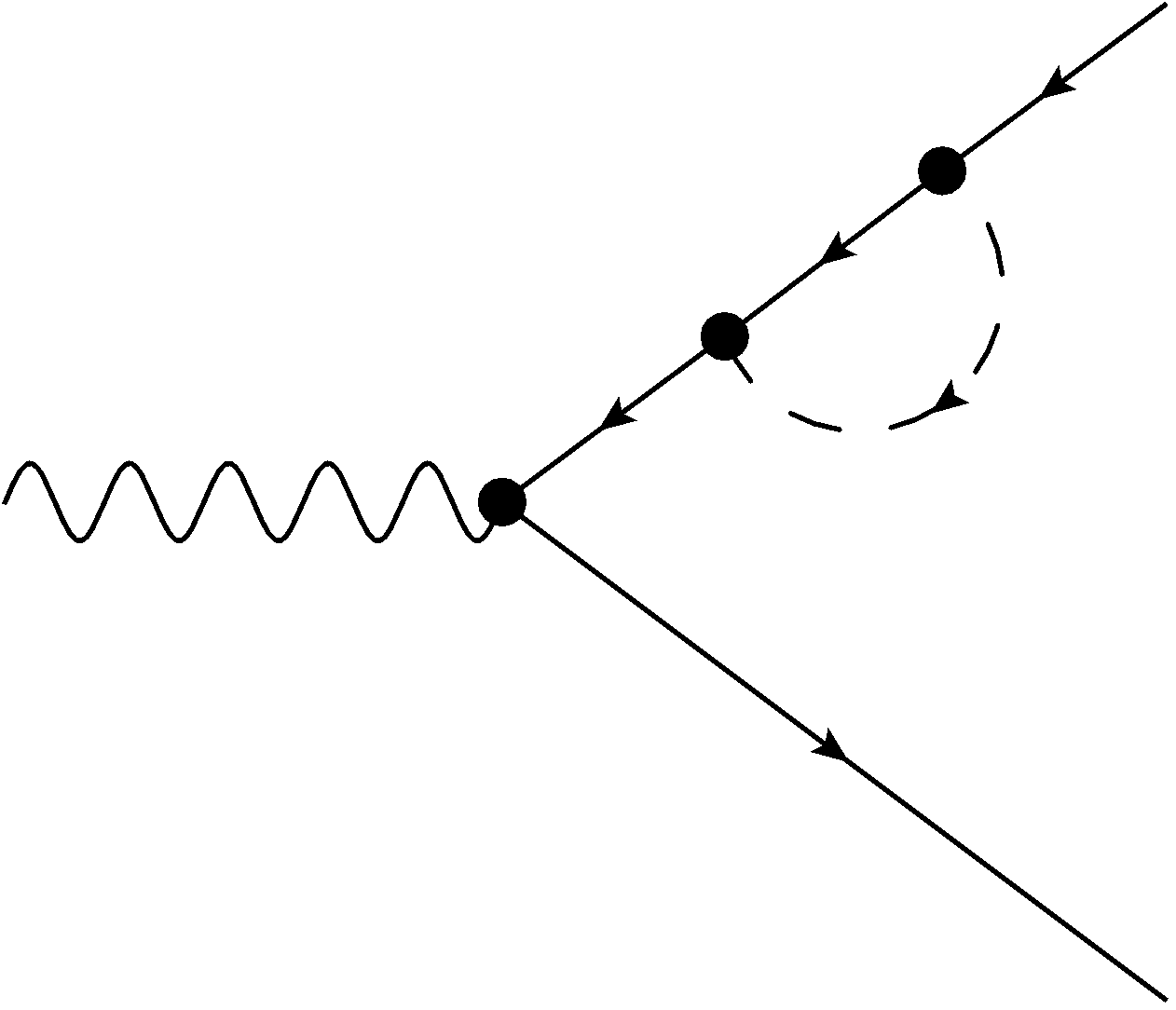} 
\put (40,80) {$(c)$}
\put (10,52) {$q$} 
\put (80,80) {$p_{1}$} \put (80,5) {$p_{3}$}
\put (75,45) {$k$}
\put (45,54) {$p_{1}$}

\put (54,66) {$p_1-k$}

\put (90,75) {$\psi_1$} \put (90,12) {$\psi_3$} 
\put (52,46) {$\psi_3$} \put (72,61) {$\psi_2$} 

\end{overpic}
\begin{overpic}[width=5.5cm]
{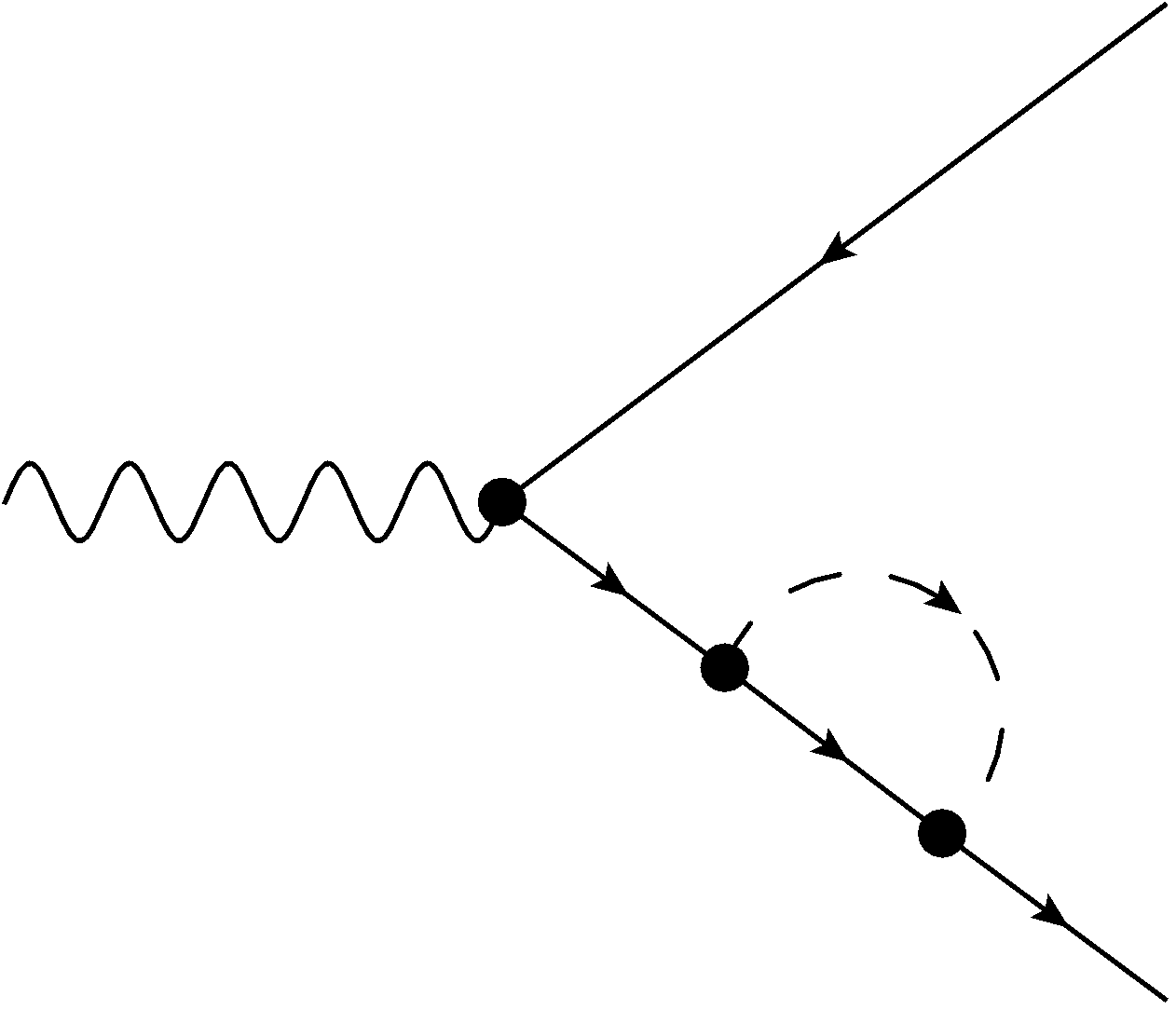} 
\put (40,80) {$(d)$}
\put (10,52) {$q$} 
\put (80,80) {$p_{1}$} \put (80,5) {$p_{3}$}
\put (75,38) {$k$}

\put (48,31) {$p_3$}
\put (55,17) {$p_3-k$}

\put (90,75) {$\psi_1$} \put (90,12) {$\psi_3$} 
\put (70,25) {$\psi_2$} \put (55,36) {$\psi_1$} 

\end{overpic}

\caption{\label{fig:triangle}Triangle diagrams that give rise to the effective
flavor-changing operator. The UV divergences of these diagrams should
cancel in the summation due to gauge invariance.}
\end{figure}

The $U(1)$ model being considered here contains a massive scalar
$\phi$ and three massless fermions $\psi_{1}$, $\psi_{2}$ and $\psi_{3}$.
They are all charged under the $U(1)$ gauge symmetry so the Lagrangian
is 
\begin{align}
{\cal L} & =\sum_{i=1}^{3}\overline{\psi}_{i}i\slashed{D}_{\mu}\psi_{i}+|D_{\mu}\phi|^{2}-m_{\phi}^{2}\phi\phi^{*} 
+\left(y_{21}\overline{\psi}_{2}\phi\psi_{1}+y_{23}\overline{\psi}_{2}\phi\psi_{3}+{\rm h.c.}\right),\label{eq:L}
\end{align}
where 
\begin{equation}
D_{\mu}=\partial_{\mu}-igQA_{\mu},\ \ Q=\begin{cases}
Q_{\phi} & {\rm for}\ \phi\\
Q_{i} & {\rm for}\ \psi_{i}
\end{cases}.\label{eq:L-1}
\end{equation}
In Eq.~(\ref{eq:L}) we have included only two Yukawa couplings $y_{12}$
and $y_{23}$ because this is the minimal requirement to obtain the
effective flavor-changing operator $A_{\mu}\overline{\psi}_{1}\gamma^{\mu}\psi_{3}$
at the 1-loop level, as indicated in Fig.~\ref{fig:triangle}. Including
other Yukawa terms ($\overline{\psi}_{1}\phi\psi_{3}$ or $\overline{\psi}_{1}\phi^{*}\psi_{3}$)
would only complicate the scenario and may not allowed by the $U(1)$
charges\footnote{For example, if $|Q_{\phi}|\neq Q_{1}-Q_{3}$, the 1-3 Yukawa mixing
terms can be forbidden by the $U(1)$ symmetry.}. Due to the $U(1)$ charge conservation, the Yukawa interactions in
Eq.~(\ref{eq:L}) are allowed when 
\begin{equation}
Q_{\phi}-Q_{2}+Q_{1}=Q_{\phi}-Q_{2}+Q_{3}=0,\label{eq:L-2}
\end{equation}
which further implies 
\[
Q_{1}=Q_{3}.
\]
There are two potential problems when applying the above $U(1)$ model
to SM extensions. First, the $Z$ boson in the SM is massive
while here the $U(1)$ gauge boson is massless if the gauge symmetry
is unbroken. To make the result in this appendix applicable to massive
gauge bosons, we manually introduce a mass $m_{A}$ for the gauge
boson.  The gauge boson mass could be generated by introduce another
scalar field with spontaneous symmetry breaking but we would rather
refrain from involving such details. 

Another problem is about the chiral fermions. The SM, as a chiral
theory of fermions, has different gauge interactions for left\textendash handed
and right-handed fermions. For example, $e_{L}$ and $e_{R}$ have
different $Z$-vertices. The $U(1)$ model considered here is parity
conserving, i.e.\ both the left-handed and right-handed components of
$\psi_{i}$ are equally coupled to the gauge and scalar bosons. This
problem can be easily solved if the Dirac spinors are decomposed into
Weyl spinors. In the $U(1)$ model, the two sets of Weyl components
$(\psi_{L1},\ \psi_{R2},\ \psi_{L3})$ and $(\psi_{R1},\ \psi_{L2},\ \psi_{R3})$
do not couple to each other directly because 
\begin{align}
\overline{\psi}_{i}i\slashed{D}_{\mu}\psi_{i} & =\overline{\psi}_{Li}i\slashed{D}_{\mu}\psi_{Li}+\overline{\psi}_{Ri}i\slashed{D}_{\mu}\psi_{Ri},\label{eq:L-3}\\
\overline{\psi}_{2}\phi\psi_{1} & =\overline{\psi}_{R2}\phi\psi_{L1}+\overline{\psi}_{L2}\phi\psi_{R1},\label{eq:L-4}\\
\overline{\psi}_{2}\phi\psi_{3} & =\overline{\psi}_{R2}\phi\psi_{L3}+\overline{\psi}_{L2}\phi\psi_{R3}.\label{eq:L-5}
\end{align}
Therefore, in the diagrams in Fig.~\ref{fig:triangle}, all the fermion
lines can also be regarded as either $(\psi_{L1},\ \psi_{R2},\ \psi_{L3})$
or $(\psi_{R1},\ \psi_{L2},\ \psi_{R3})$. At the end of this section,
we will also present the result for a chiral $U(1)$.

It is important to notice that the sum of all the four diagrams in
Fig.~\ref{fig:triangle} is finite, as pointed out in Ref.~\cite{Bilenky:1993bt}.
The UV divergences necessarily cancel out if the model is renormalizable,
otherwise there is no corresponding counter term to cancel the infinity.
We will show the cancellation explicitly in the following calculation.

Now let us compute the four diagrams (a)-(d) in Fig.~\ref{fig:triangle}.
The relevant interactions are 
\begin{equation}
{\cal L}\supset y_{21}\overline{\psi}_{2}\phi\psi_{1}+y_{23}^{*}\overline{\psi}_{3}\phi^{*}\psi_{2}+g\sum_{i}Q_{i}A_{\mu}\overline{\psi}_{i}\gamma^{\mu}\psi_{i}+igQ_{\phi}A^{\mu}\phi^{*}\overleftrightarrow{\partial}_{\mu}\phi,\label{eq:L-6}
\end{equation}
where $\phi^{*}\overleftrightarrow{\partial}_{\mu}\phi\equiv\phi^{*}\partial_{\mu}\phi-\phi\partial_{\mu}\phi^{*}$.
Then it is straightforward to write down the amplitudes 
\begin{equation}
i{\cal M}_{a}=\int\frac{d^{4}k}{(2\pi)^{4}}\overline{u}(p_{3})iy_{23}^{*}\frac{i}{(p_{3}-k)\cdot\gamma}igQ_{2}\gamma^{\mu}\epsilon_{\mu}(q)\frac{i}{(p_{1}-k)\cdot\gamma}iy_{21}\frac{i}{k^{2}-m_{\phi}^{2}}u(p_{1}),\label{eq:L-7}
\end{equation}
\begin{equation}
i{\cal M}_{b}=\int\frac{d^{4}k}{(2\pi)^{4}}\overline{u}(p_{3})iy_{23}^{*}\frac{i}{\slashed{k}}iy_{21}\frac{i}{(p_{3}-k)^{2}-m_{\phi}^{2}}\left[-igQ_{\phi}(p_{1}+p_{3}-2k)^{\mu}\right]\epsilon_{\mu}(q)\frac{i}{(p_{1}-k)^{2}-m_{\phi}^{2}}u(p_{1}),\label{eq:L-8}
\end{equation}
\begin{equation}
i{\cal M}_{c}=\int\frac{d^{4}k}{(2\pi)^{4}}\overline{u}(p_{3})igQ_{3}\gamma^{\mu}\epsilon_{\mu}(q)\frac{i}{\slashed{p}_{1}-m_{3}}iy_{23}^{*}\frac{i}{(p_{1}-k)\cdot\gamma}iy_{21}\frac{i}{k^{2}-m_{\phi}^{2}}u(p_{1}),\label{eq:L-9}
\end{equation}
\begin{equation}
i{\cal M}_{d}=\int\frac{d^{4}k}{(2\pi)^{4}}\overline{u}(p_{3})iy_{23}^{*}\frac{i}{(p_{3}-k)\cdot\gamma}\frac{i}{k^{2}-m_{\phi}^{2}}iy_{21}\frac{i}{\slashed{p}_{3}-m_{1}}igQ_{1}\gamma^{\mu}\epsilon_{\mu}(q)u(p_{1}).\label{eq:L-10}
\end{equation}
Here we assume that the fermions $\psi_{1,\thinspace3}$ have very
small masses $m_{1,\thinspace3}$ so that the standard technique of
extracting form factors can be applied. After that, we will take the
limit $m_{1, 3}\rightarrow0$. For simplicity, we evaluate
the amplitudes with all external momenta on shell  so that amplitudes
can be organized by three form factors $F_{1}$, $F_{2}$ and $F_{3}$
as follows
\begin{equation}
i{\cal M}_{a}+i{\cal M}_{b}+i{\cal M}_{c}+i{\cal M}_{d}=\overline{u}(p_{3})\left[\left(\gamma^{\mu}-\frac{\slashed{q}q^{\mu}}{q^{2}}\right)F_{1}(q^{2})+\frac{i\sigma^{\mu\nu}q_{\nu}}{m_{1}+m_{3}}F_{2}(q^{2})+\frac{2q^{\mu}}{m_{1}+m_{3}}F_{3}(q^{2})\right]u(p_{1})\epsilon_{\mu}(q).\label{eq:L-13}
\end{equation}
Since $\slashed{q}=\slashed{p}_{3}-\slashed{p}_{1}$ and $\slashed{p}_{1}u(p_{1})=m_{1}$,
$\overline{u}(p_{3})\slashed{p}_{3}=m_{3}$, the $F_{1}$ term actually
reduces to $\gamma^{\mu}F_{1}(q^{2})$ in the zero mass limit ($m_{1,3}\rightarrow0$).
We use the computer program {\tt Package-X} \cite{Patel:2015tea}
to compute the loop integrals in dimensional regularization. The form
factors can be directly extracted by using the corresponding projectors
in {\tt Package-X}. In this paper, we are only interested in the
$F_{1}$ form factor. Let us first check the UV divergence in $F_{1}$.
Since we are using dimensional regularization, the UV divergent part
is proportional to $1/\epsilon\equiv2/(d-4)$:
\begin{equation}
F_{1}^{({\rm divergent})}=\frac{gy_{23}^{*}y_{21}}{2\epsilon}\left(Q_{2}-Q_{\phi}+\frac{-Q_{3}m_{1}}{m_{1}-m_{3}}+\frac{Q_{1}m_{3}}{m_{1}-m_{3}}\right),\label{eq:L-14}
\end{equation}
where the terms proportional to $Q_{2}$, $Q_{\phi}$, $Q_{3}$, and
$Q_{1}$ correspond to the contributions of diagrams ($a$), ($b$),
($c$), and ($d$), respectively (note that each of these diagrams
has a distinct gauge interaction vertex and a characteristic $U(1)$
charge). Eq.~(\ref{eq:L-14}) can also be written as 
\begin{equation}
F_{1}^{({\rm divergent})}=\frac{gy_{23}^{*}y_{21}}{2\epsilon}\frac{m_{3}(Q_{\phi}-Q_{2}+Q_{1})-m_{1}(Q_{\phi}-Q_{2}+Q_{3})}{m_{1}-m_{3}},\label{eq:L-15}
\end{equation}
which, according to Eq.~(\ref{eq:L-2}), implies that the UV divergence
vanishes if the $U(1)$ charges are conserved.

Taking $Q_{3}=Q_{1}$, $Q_{\phi}=Q_{2}-Q_{1}$ and $(m_{1},\ m_{3})\rightarrow0$,
the finite part of $F_{1}$ is
\begin{equation}
F_{1}^{({\rm finite})}=\frac{igy_{23}^{*}y_{21}\left(f(r)Q_{1}+h(r)Q_{2}\right)}{16\pi^{2}},\label{eq:L-16}
\end{equation}
with
\begin{equation}
r\equiv\frac{m_{A}^{2}}{m_{\phi}^{2}},\label{eq:L-17}
\end{equation}
\begin{equation}
\omega\equiv-r-i0^{+},\label{eq:L-81}
\end{equation}
\begin{equation}
f(r)=\frac{1}{4\omega}\left[-4C_{101}(\omega)+2(\omega+2)B_{0\Lambda}(\omega)+5\omega+4\right],\label{eq:L-18}
\end{equation}
\begin{equation}
h(r)=\frac{1}{4\omega}\left[4C_{010}(\omega)+4C_{101}(\omega)+2(\omega+2)\left(\log\frac{1}{\omega}-B_{0\Lambda}(\omega)\right)-4\omega\right].\label{eq:L-19}
\end{equation}
Here $B_{0\Lambda}$, $C_{101}$, $C_{010}$ are parts of the scalar
Passarino-Veltman functions, with the explicit forms given below:
\begin{equation}
B_{0\Lambda}(\omega)=-\frac{1}{\omega}\sqrt{\omega(\omega+4)}\log\left(\frac{\omega+\sqrt{\omega(\omega+4)}+2}{2}\right),\label{eq:L-20}
\end{equation}
\begin{align}
C_{101}(\omega) & =\frac{\pi^{2}}{6\omega}+\frac{1}{2\omega}\left[\log^{2}\left(\frac{\omega-\sqrt{\omega(\omega+4)}}{2\omega}\right)-\log^{2}\left(\frac{\omega+\sqrt{\omega(\omega+4)}+2}{\sqrt{\omega(\omega+4)}-\omega}\right)\right]\nonumber \\
 & -\frac{1}{\omega}\text{Li}_{2}(\omega+1)-\frac{1}{\omega}\text{Li}_{2}\left(\frac{2(\omega+1)}{\omega-\sqrt{\omega(\omega+4)}}\right)+\frac{1}{\omega}\text{Li}_{2}\left(\frac{2}{\sqrt{\omega(\omega+4)}-\omega}\right)\nonumber \\
 & -\frac{1}{\omega}\text{Li}_{2}\left(\frac{2}{\omega+\sqrt{\omega(\omega+4)}+2}\right)+\frac{1}{\omega}\text{Li}_{2}\left(\frac{1}{2}(\omega+1)\left(\omega+\sqrt{\omega(\omega+4)}+2\right)\right),\label{eq:L-21}
\end{align}
\begin{equation}
C_{010}(\omega)=-\frac{6\text{Li}_{2}\left(\frac{\omega-1}{\omega}\right)+3\log^{2}\left(\frac{1}{\omega}\right)+\pi^{2}}{6\omega}.\label{eq:L-22}
\end{equation}
Using the identities of the dilogarithm function (\ref{eq:L-23})-(\ref{eq:L-25}),
we can make a series expansion in $r$ and obtain Eq.~(\ref{eq:L-59})
and Eq.~(\ref{eq:L-60}).

In summary, the triangle diagrams can generate the following effective
vector vertex
\begin{equation}
{\cal L}_{{\rm eff}}=g_{31}^{(1)}A_{\mu}\overline{\psi}_{3}\gamma^{\mu}\psi_{1},\label{eq:L-12}
\end{equation}
where 
\begin{equation}
g_{31}^{(1)}=\frac{gy_{23}^{*}y_{21}}{16\pi^{2}}\left(f(r)Q_{1}+h(r)Q_{2}\right).\label{eq:L-11}
\end{equation}
Note that the result is applicable only when the $U(1)$ charges are
conserved\textemdash see Eq.~(\ref{eq:L-2}).

For a chiral $U(1)$ theory, we can still use the above result by
simply replacing the Dirac spinors with the chiral spinors. For example,
if  only $(\psi_{L1},\ \psi_{R2},\ \psi_{L3})$ are present in the
model, then we have 
\begin{equation}
{\cal L}_{{\rm eff}}=g_{31}^{(1)}A_{\mu}\overline{\psi}_{L3}\gamma^{\mu}\psi_{L1}=g_{31}^{(1)}A_{\mu}\overline{\psi}_{3}\gamma^{\mu}P_{L}\psi_{1},\label{eq:L-12-1}
\end{equation}
where $g_{31}^{(1)}$ is the same as Eq.~(\ref{eq:L-11}), and $(Q_{1},\ Q_{2},\ Q_{3})$
should be the $U(1)$ charges of $(\psi_{L1},\ \psi_{R2},\ \psi_{L3})$
respectively.

\section{The box diagram\label{sec:box-diagram}}

\begin{figure}
\centering

\begin{overpic}[width=5.5cm]
{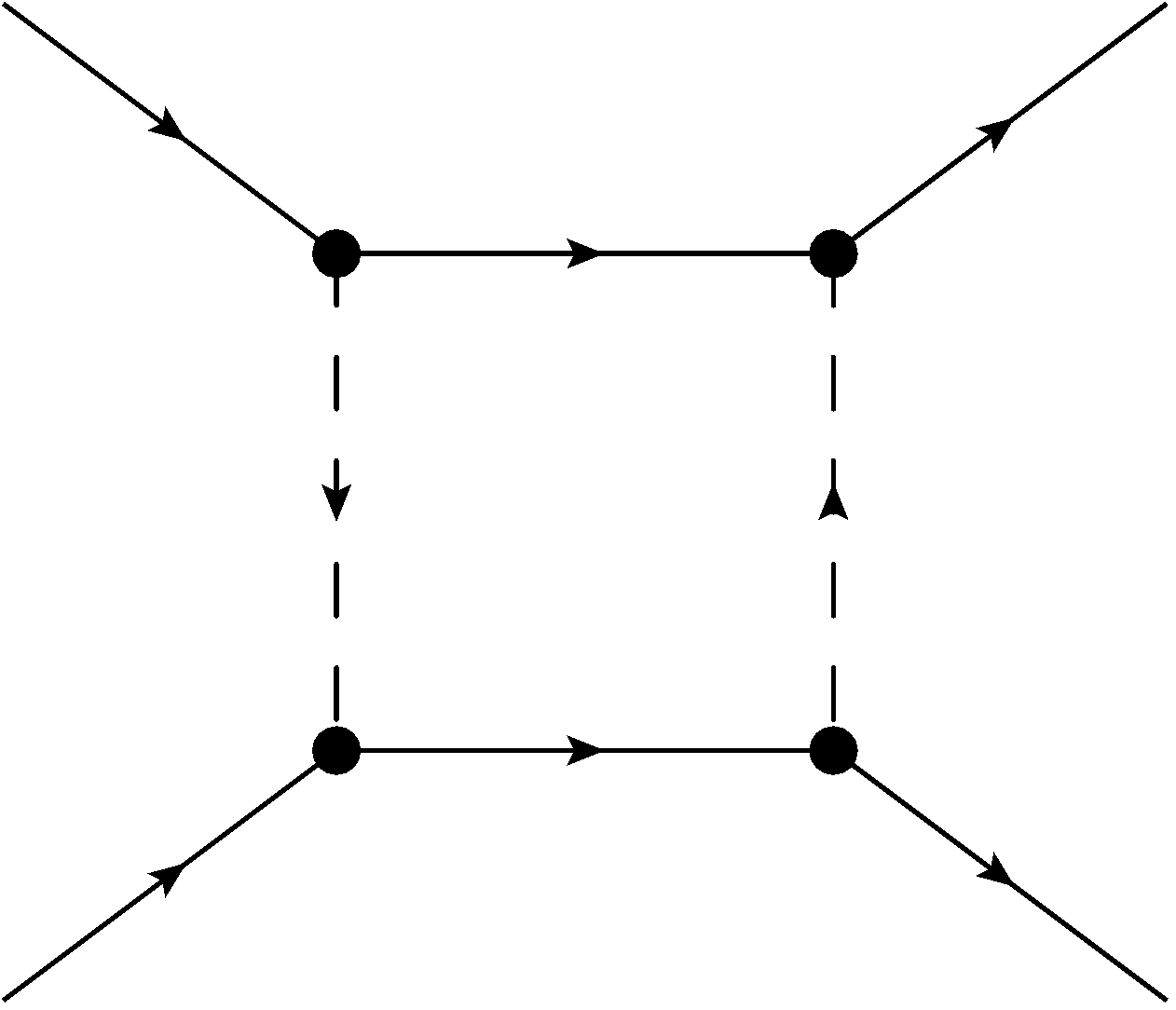} 
\put (15,77) {$\psi_{1}$}
\put (48,68) {$\psi_{2}$}
\put (80,77) {$\psi_{3}$}
\put (10,5) {$\psi_{4}$}
\put (48,17) {$\psi_{5}$}
\put (80,6) {$\psi_{6}$}

\put (48,57) {$k_2$}
\put (48,25) {$k_5$}
\put (31,45) {$k_L$}
\put (62,45) {$k_R$}

\put (10,71) {$p_{1}$}
\put (87,71) {$p_{3}$}
\put (10,15) {$p_{4}$}
\put (90,11) {$p_{6}$}

\end{overpic}

\caption{\label{fig:box}Box diagram that generates the effective four-fermion
operator $\overline{\psi}_{3}\gamma^{\mu}\psi_{1}\overline{\psi}_{6}\gamma_{\mu}\psi_{4}$. }
\end{figure}
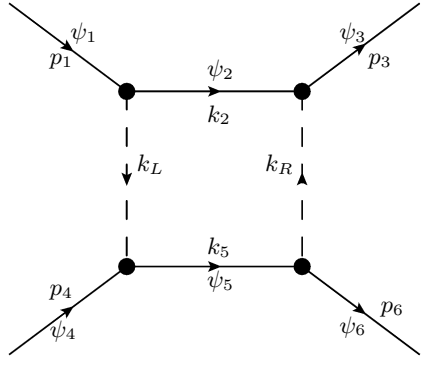
This appendix computes the box diagram in a general model with six
fermions $\psi_{i}$ ($i=1, 2,\cdots,6$) and one complex scalar field
$\phi$, with the following Yukawa interactions: 
\begin{align}
{\cal L}\supset & y_{21}\overline{\psi}_{2}\phi\psi_{1}+y_{23}\overline{\psi}_{2}\phi\psi_{3}+y_{45}\overline{\psi}_{4}\phi\psi_{5}+y_{65}\overline{\psi}_{6}\phi\psi_{5}\nonumber \\
+ & y_{21}^{*}\overline{\psi}_{1}\phi^{*}\psi_{2}+y_{23}^{*}\overline{\psi}_{3}\phi^{*}\psi_{2}+y_{45}^{*}\overline{\psi}_{5}\phi^{*}\psi_{4}+y_{65}^{*}\overline{\psi}_{5}\phi^{*}\psi_{6}.\label{eq:L-26}
\end{align}
The second line is just the hermitian conjugate of the first line.
For convenience of later use, we write the hermitian conjugate terms
explicitly. 

The box diagram we will compute is shown in Fig.~\ref{fig:box},
according to which we can straightforwardly write down the amplitude
\begin{equation}
i{\cal M}_{{\rm box}}=\int\frac{d^{4}k}{(2\pi)^{4}}\overline{u}(p_{3})iy_{23}^{*}\frac{i}{\slashed{k}_{2}}iy_{21}u(p_{1})\frac{i}{k_{L}^{2}-m_{\phi}^{2}}\frac{i}{k_{R}^{2}-m_{\phi}^{2}}\overline{u}(p_{6})iy_{65}\frac{i}{\slashed{k}_{5}}iy_{45}^{*}u(p_{4}).\label{eq:L-27}
\end{equation}
where  the fermions are all massless and the scalar has mass $m_{\phi}^{2}$. 

The amplitude is finite and can be computed directly. Since we are
interested in the heavy scalar mass limit, let us take the 
zero external momentum limit $(p_{1},\ p_{3},\ p_{4},\ p_{6})/m_{\phi}\rightarrow0$:
\begin{equation}
i{\cal M}_{{\rm box}}=\overline{u}(p_{3})iy_{23}^{*}\gamma^{\mu}iy_{21}u(p_{1})\overline{u}(p_{6})iy_{65}\gamma^{\nu}iy_{45}^{*}u(p_{4})I(m_{\phi}^{2},\ m_{\phi}^{2}),\label{eq:L-28}
\end{equation}
where we define the integral
\begin{equation}
I(m_{a}^{2},\ m_{b}^{2})=\int\frac{d^{4}k}{(2\pi)^{4}}\frac{i(-k^{\mu})}{k^{2}}\frac{i(k^{\nu})}{k^{2}}\frac{i}{k^{2}-m_{a}^{2}}\frac{i}{k^{2}-m_{b}^{2}}.\label{eq:L-29}
\end{equation}
It can be evaluated straightforwardly:
\begin{equation}
I(m_{a}^{2},\ m_{b}^{2})=\frac{i}{16\pi^{2}}\frac{\log\left(\frac{m_{a}^{2}}{m_{b}^{2}}\right)g^{\mu\nu}}{4\left(m_{a}^{2}-m_{b}^{2}\right)}.\label{eq:L-30}
\end{equation}
In the equal mass limit ($m_{a}^{2}=m_{b}^{2}=m_{\phi}^{2}$), it is
\begin{equation}
I(m_{\phi}^{2},\ m_{\phi}^{2})=\frac{i}{16\pi^{2}}\frac{1}{4m_{\phi}^{2}}g^{\mu\nu}.\label{eq:L-31}
\end{equation}

In summary, the box diagram generates the following four-fermion effect
operator 
\begin{align}
{\cal L}_{{\rm eff}} & =\frac{1}{16\pi^{2}}\frac{y_{23}^{*}y_{21}y_{65}y_{45}^{*}}{4m_{\phi}^{2}}\overline{\psi}_{3}\gamma^{\mu}\psi_{1}\overline{\psi}_{6}\gamma_{\mu}\psi_{4}\label{eq:L-32}\\
 & =\frac{1}{16\pi^{2}}\frac{y_{23}^{*}y_{21}y_{65}y_{45}^{*}}{4m_{\phi}^{2}}\overline{\psi}_{3}\gamma^{\mu}\psi_{1}\overline{\psi_{4}^{c}}(-\gamma_{\mu})\psi_{6}^{c}.\label{eq:L-32-1}
\end{align}

\section{\label{sec:appC}Some useful identities and transformations }

In this work, we need to frequently transform Dirac matrices and 
spinor products from one to another. Besides, in the loop calculation,
we also need some useful identities about the dilogarithm functions.
Therefore, we compile them in this appendix.

The left- and right-handed projectors are defined as
\begin{equation}
P_{L}\equiv\frac{1-\gamma^{5}}{2},\ P_{R}\equiv\frac{1+\gamma^{5}}{2}.\label{eq:L-69}
\end{equation}
Products of $P_{L/R}$ with the Dirac matrices can be transformed
using 
\begin{equation}
\gamma^{5}P_{L}=P_{L}\gamma^{5}=-P_{L},\ P_{R}\gamma^{5}=\gamma^{5}P_{R}=P_{R},\label{eq:L-71}
\end{equation}
\begin{equation}
\gamma_{\mu}\gamma^{5}=-\gamma^{5}\gamma_{\mu},\ \gamma^{\mu}P_{L}=P_{R}\gamma^{\mu},\ \ \gamma^{\mu}P_{R}=P_{L}\gamma^{\mu}.\label{eq:L-72}
\end{equation}
Defining
\begin{equation}
\sigma_{\mu\nu}\equiv\frac{i}{2}[\gamma^{\mu},\ \gamma^{\nu}],\label{eq:L-70}
\end{equation}
we also have
\[
P_{L}\sigma_{\mu\nu}=\sigma_{\mu\nu}P_{L},\ P_{R}\sigma_{\mu\nu}=\sigma_{\mu\nu}P_{R}.
\]
The left- and right-handed components of a Dirac spinor $\psi$ are
defined as
\begin{equation}
\psi_{L}\equiv P_{L}\psi,\ \ \psi_{R}\equiv P_{R}\psi.\label{eq:L-73}
\end{equation}
The charge conjugate of $\psi$ is defined as 
\begin{equation}
\psi^{c}\equiv-i\gamma^{2}\psi^{*}.\label{eq:L-74}
\end{equation}
The left- and right-handed projections and the charge conjugation
are related by 
\begin{equation}
\overline{\psi_{L}}=\overline{\psi}P_{R},\ \overline{\psi_{R}}=\overline{\psi}P_{L},\label{eq:L-75}
\end{equation}
\begin{equation}
\psi_{L}^{c}\equiv(\psi_{L})^{c}=-i\gamma^{2}P_{L}\psi^{*}=P_{R}\psi^{c}.\label{eq:L-76}
\end{equation}
For two different Dirac spinors $\psi_{1}$ and $\psi_{2}$, we have
\begin{equation}
\overline{\psi_{1}^{c}}\psi_{2}=\overline{\psi_{2}^{c}}\psi_{1},\ \overline{\psi_{1}}\psi_{2}^{c}=\overline{\psi_{2}}\psi_{1}^{c},\ \overline{\psi_{1}^{c}}\psi_{2}^{c}=\overline{\psi_{2}}\psi_{1},\label{eq:L-77}
\end{equation}
\begin{equation}
\overline{\psi_{1}^{c}}\gamma^{\mu}\psi_{2}^{c}=-\overline{\psi_{2}}\gamma^{\mu}\psi_{1}.\label{eq:L-78}
\end{equation}

Turning to loop-functions needed in this study, the dilogarithm 
$\text{Li}_{2}(z)$ can be defined by 
\begin{equation}
\text{Li}_{2}(z)=\sum_{k=1}^{\infty}\frac{z^{k}}{k^{2}}=\int_{z}^{0}\frac{\log(1-t)}{t}dt.\label{eq:L-79}
\end{equation}
It has a branch cut at $z>1$, so in many cases we need the following
identities 
\begin{equation}
\text{Li}_{2}(\frac{1}{z})=-\text{Li}_{2}(z)-\frac{1}{2}\log^{2}(-z)-\frac{\pi^{2}}{6},\label{eq:L-23}
\end{equation}
\begin{equation}
\text{Li}_{2}(1-\frac{1}{z})=\text{Li}_{2}(z)-\frac{1}{2}\log^{2}(z)+\log(1-z)\log(z)-\frac{\pi^{2}}{6},\label{eq:L-24}
\end{equation}
\begin{equation}
\text{Li}_{2}(1-z)=-\text{Li}_{2}(z)-\log(1-z)\log(z)+\frac{\pi^{2}}{6},\label{eq:L-25}
\end{equation}
to transform some dilogarithmic singularities to logarithmic singularities
which are easier to handle.

\bibliographystyle{apsrev4-1}
\bibliography{ref}

\end{document}